\documentclass[apjl]{emulateapj}
\usepackage{psfig,amsfonts,amsmath,graphicx,natbib,apjfonts,lscape}

\def\nod{\nodata}
\def\cfa{1}

\begin{document}

\title{A Beaming-Independent Estimate of the Energy Distribution of
Long Gamma-Ray Bursts: Initial Results and Future Prospects}

\author{ 
I.~Shivvers\altaffilmark{\cfa} \&
E.~Berger\altaffilmark{\cfa}
}

\altaffiltext{\cfa}{Harvard-Smithsonian Center for Astrophysics, 60
Garden Street, Cambridge, MA 02138}

\begin{abstract} We present single-epoch radio afterglow observations
of $24$ long-duration gamma-ray burst (GRB) on a timescale of $\gtrsim
100$ d after the burst.  These observations trace the afterglow
evolution when the blastwave has decelerated to mildly- or
non-relativistic velocities and has roughly isotropized.  We infer
beaming-independent kinetic energies using the Sedov-Taylor
self-similar solution, and find a median value for the sample of
detected bursts of about $7\times 10^{51}$ erg, with a $90\%$
confidence range of $1.1\times 10^{50}-3.3\times 10^{53}$ erg.  Both
the median and $90\%$ confidence range are somewhat larger than the
results of multi-wavelength, multi-epoch afterglow modeling (including
large beaming corrections), and the distribution of beaming-corrected
$\gamma$-ray energies.  This is due to bursts in our sample with only
a single-frequency observation for which we can only determine an
upper bound on the peak of the synchrotron spectrum.  This limitation
leads to a wider range of allowed energies than for bursts with a
well-measured spectral peak.  Our study indicates that single-epoch
centimeter-band observations covering the spectral peak on a timescale
of $\delta t\sim 1$ yr can provide a robust estimate of the total
kinetic energy distribution with a small investment of telescope time.
The substantial increase in bandwidth of the EVLA (up to 8 GHz
simultaneously with full coverage at $1-40$ GHz) will provide the
opportunity to estimate the kinetic energy distribution of GRBs with
only a few hours of data per burst.  \end{abstract}

\keywords{gamma-rays:bursts}

\section{Introduction}
\label{sec:intro}

The energy budget of gamma-ray bursts (GRBs) provides fundamental
insight into the nature of the explosions, the resulting ejecta
properties, and the identity of the central compact remnant
(``engine'').  While the isotropic-equivalent $\gamma$-ray energy
($E_{\rm\gamma,iso}$) can be easily determined from a measurement of
the burst fluence and redshift, a complete accounting of the energy
budget requires detailed observations of the afterglow emission.  The
afterglow observations provide a measure of the isotropic-equivalent
blastwave kinetic energy ($E_{\rm K,iso}$), as well as the explosion
geometry (quantified by a jet opening angle, $\theta_j$).  The
resulting beaming corrections, $f_b^{-1}\equiv 1-{\rm cos}(\theta_j)$,
can be substantial, approaching three orders of magnitude in some
cases \citep{fks+01,pk02,bkf03,bfk03}.  To properly determine $E_{\rm
K,iso}$ and $f_b$ it is essential to observe the afterglows from radio
to X-rays over timescales of hours to weeks, clearly a challenging
task.  This is particularly a problem for the subset of ``dark'' GRBs
for which the lack of detected optical emission, or large extinction,
prevent a determination of $E_{\rm K,iso}$ and likely $f_b$ (e.g.,
\citealt{bkb+02,pfg+02}).

Over the past decade detailed afterglow observations have been
obtained at a great cost of telescope time for about $20$
long-duration GRBs, with the basic result that the beaming corrections
are large and diverse, leading to typical true energies of
$E_\gamma\sim E_K\sim 10^{51}$ erg
\citep{fks+01,pk01,pk02,bkf03,bkp+03,bfk03}.  More recently, it has
been recognized that some nearby long GRBs have much lower energies,
$E_{\rm iso}\sim 10^{49}-10^{50}$ erg, and appear to be
quasi-isotropic \citep{kfw+98,skb+04,skn+06}.  Similarly, some bursts
appear to have large beaming-corrected energies of $\sim 10^{52}$ erg
\citep{cfh+10,cfh+10b}.  The existence of these highly energetic
bursts depends at least in part on the ability to correctly infer
their large beaming corrections.  Indeed, the inference of jet opening
angles from breaks in the afterglow light curves has become
controversial in recent years due to conflicting trends in optical and
X-ray light curves \citep{lrz+08,rlb+09}.  Similarly, in some cases a
two-component jet has been inferred, with a narrow core dominating the
$\gamma$-ray emission and a wider component dominating the afterglow
emission \citep{bkp+03,rks+08}.  Numerical simulations suggest that
off-axis viewing angles can also lead to shallow breaks that may be
missed or mis-interpreted \citep{vzm10}.

In addition to potential difficulties with the inference of $f_b$, the
$\gamma$-ray and kinetic energies measured from the early afterglow
emission only pertain to the relativistic ejecta.  The existence of a
substantial component of mildly relativistic ejecta can only be
determined from observations at late times when such putative material
can refresh the forward shock.  Clearly, the existence of substantial
energy in a slow ejecta component will place crucial constraints on
the activity lifetime of the central engine.

Such late-time observations also have the added advantage that they
probe the blastwave when it has decelerated to non-relativistic
velocities and hence roughly approaches isotropy \citep{fwk00,lw00}.
This allows us to use the well-established Sedov-Taylor self-similar
solution, with negligible beaming corrections, to estimate the total
kinetic energy of both the decelerated ejecta and any additional
initially non-relativistic material.  Since the peak of the afterglow
spectrum on these timescales is located in the radio band, the lack of
optical afterglow emission (e.g., due to extinction) does not have an
effect on the ability to determine $E_K$.

This approach was first exploited by \citet{fwk00} to model the
late-time radio afterglow emission of GRB\,970508 (at $\delta t\gtrsim
100$ d) from which the kinetic energy was inferred to be $E_K\sim
5\times 10^{50}$ erg.  \citet{bkf04} used the same approach to model
the radio afterglow emission of GRB\,980703 on timescales of $\gtrsim
40$ d, and to re-model GRB\,970508.  They found kinetic energies of
$E_K\sim 3\times 10^{51}$ erg for both bursts.  Finally,
\citet{fsk+05} modeled the radio emission from GRB\,030329 at $\delta
t\gtrsim 50$ d and found $E_K\sim 10^{51}$ erg.  Only 3 bursts have
been studied in this fashion so far because only those events have
well-sampled radio light curves on the relevant timescales of $\delta
t\gtrsim 100$ d.

However, the kinetic energy can still be estimated using the same
methodology even from fragmentary late-time radio observations.  Such
an approach will naturally result in larger uncertainties for each
burst, but it can be applied to a much larger sample of events.  Here
we present such an analysis for $24$ long-duration GRBs with radio
observations at $\gtrsim 100$ d, but with only $1-3$ data points (at
1.4 to 8.5 GHz) per burst.  Using these observations we infer robust
ranges for the kinetic energy of each burst and for the population as
a whole.  The plan of the paper is as follows.  The radio observations
are summarized in \S\ref{sec:obs}.  The model for synchrotron emission
from a Sedov-Taylor blastwave, and the various assumptions we employ
are presented in \S\ref{sec:model}.  In \S\ref{sec:res} we detail the
resulting kinetic energies and the range for the overall sample, and
we compare these results to multi-wavelength analyses of early
afterglows in \S\ref{sec:comp}.  We conclude with a discussion of
future prospects.

\section{Radio Data}
\label{sec:obs}

We use radio observations of $24$ long GRBs at $\delta t\gtrsim 100$ d
since on those timescales the blastwave is expected to become
non-relativistic and roughly isotropic \citep{lw00}, and the peak of
the afterglow emission is at or below the centimeter band.  This has
been confirmed with detailed data in the case of GRBs 970508, 980703,
and 030329 \citep{fwk00,bkf04,fsk+05}.  We restrict the analysis to
GRBs with a known redshift and with early-time detections, which for
the case of a single detection or upper limit allow us to infer that
the peak of the spectrum has transitioned below our observing
frequency.

The observations are primarily from the Very Large Array
(VLA\footnotemark\footnotetext{The National Radio Astronomy
Observatory is a facility of the National Science Foundation operated
under cooperative agreement by Associated Universities, Inc.}), with
the exception of GRBs 980425 and 011121 which were observed with the
Australia Telescope Compact Array (ATCA).  The data were obtained
between 1997 and 2009 as part of a long-term GRB radio program (e.g.,
\citealt{fkb+03}).

For the purpose of our analysis, we separate the bursts into three
categories based on the quality of the data.  In Group A are 3 bursts
with late-time detections at multiple frequencies that constrain the
peak of the synchrotron spectrum (the same three bursts that have been
studied in detail by \citealt{fwk00,bkf04,fsk+05}).  In Group B are 11
bursts with single-frequency detections, while Group C consists of 10
GRBs with late-time non-detections.  The VLA measurements and relevant
burst properties are listed in table~\ref{tab:data}.

\section{Synchrotron Emission from a Non-Relativistic Blastwave}
\label{sec:model}

Our modeling of the radio data follows the methodology of
\citet{fwk00} and \citet{bkf04} for the case of a uniform density
medium\footnotemark\footnotetext{Since we use a single epoch of
observations for each GRB our inferred density can be easily converted
to a mass loss rate for the case of a wind medium.  The difference in
dynamical evolution between these two models does not have an effect
in this case.}.  For the typical expected parameters of long GRBs, the
initially collimated blastwave approaches spherical symmetry and
decelerates to non-relativistic velocity on similar timescales,
$t_s\approx 150(E_{\rm K,iso,52}/n_e)^{1/4} t_{\rm j,d}^{1/4}$ d and
$t_{NR}\approx 40(E_{\rm K,iso,52}/n_e)^{1/4}t_{\rm j,d}^{1/4}$ d,
respectively \citep{lw00}; here, $n_e$ is the circumburst density in
units of cm$^{-3}$ and $t_j$ is the ``jet break'' time at which the
jet begins to expand sideways (i.e., $\Gamma(t_j)\sim\theta_j^{-1}$,
where $\Gamma$ is the bulk Lorentz factor).  In this paper we assume
that the blastwave has transitioned to the non-relativistic isotropic
phase by the time of our observations and subsequently check for
self-consistency.

The blastwave dynamics in the non-relativistic phase are described by
the Sedov-Taylor self-similar solution with $r(t)\propto (E_{\rm ST}
t^2/n)^{1/5}$.  To calculate the synchrotron emission emerging from
the shock-heated material, we make the usual assumptions: (i) the
electrons are accelerated to a power-law energy distribution,
$N(\gamma)\propto\gamma^{-p}$ for $\gamma>\gamma_m$, where $\gamma_m$
is the minimum Lorentz factor; (ii) the value of $p$ is 2.2 as
inferred from several bursts (e.g., \citealt{pk01,pk02,yhs+03}); and
(iii) the energy densities in the magnetic field and electrons are
constant fractions ($\epsilon_B$ and $\epsilon_e$, respectively) of
the shock energy density.  Accounting for synchrotron emissivity and
self-absorption, and including the appropriate redshift
transformations, the flux observed at frequency $\nu$ and time $t$ is
given by \citep{fwk00,bkf04}:
\begin{equation}
F_\nu = F_0(t/t_0)^{\alpha_F}[(1+z)\nu]^{5/2}(1-e^{-\tau_\nu})
f_3(\nu/\nu_m)f^{-1}_2(\nu/\nu_m), 
\end{equation}
where the optical depth is given by:
\begin{equation}
\tau_\nu = \tau_0(t/t_0)^{\alpha_{\tau}}[(1+z)\nu]^{-(p+4)/2}
f_2(\nu/\nu_m),
\end{equation}
the synchrotron peak frequency, corresponding to electrons with
$\gamma=\gamma_m$, is given by:
\begin{equation}
\nu_m = \nu_0(t/t_0)^{\alpha_m}(1+z)^{-1},
\end{equation}
and the function $f_l(x)$ is given by
\begin{equation}
f_l(x)=\int^x_0 F(y)y^{(p-l)/2}\,dy,
\end{equation}
where $F(y)$ is an integration over Bessel functions \citep{rl79}.
The temporal indices in the case of a uniform density medium are
$\alpha_F=11/10$, $\alpha_\tau=1-3p/2$, and $\alpha_m=-3$.  The
normalizations are such that $F_0$ and $\tau_0$ are the flux density
and optical depth at a frequency of $\nu=1$ Hz at $t=t_0$, and $\nu_0$
is the synchrotron peak frequency in the rest frame of burst at
$t=t_0$.  Furthermore, the synchrotron self-absorption frequency,
$\nu_a$, is defined by the condition $\tau_\nu(\nu_a)=1$.

We fit this synchrotron model to our radio data using $F_0$, $\tau_0$,
and $\nu_0$ as free parameters.  Since we have no {\it a priori}
knowledge about the expected values of the synchrotron spectrum
parameters we assume that they follow a flat distribution in
log-space.  We note that any further assumption about the distribution
of these parameters will only serve to restrict the resulting energy
distributions, and we therefore consider our assumed flat distribution
to be conservative.  For the detected objects (Groups A and B) we
retain all solutions that reproduce the measured flux density within
the error bars, while for the non-detections we use $3\sigma$ as an
upper bound.  Moreover, for the bursts in Group A, $\nu_0$ is
constrained by the multi-frequency observations and no additional
constraints are required.  However, for the bursts in Groups B and C,
which have only a single-frequency observation, we require that both
$\nu_m$ and $\nu_a$ have values below the observing frequency since
the light curves are always declining at the time of our
observations\footnotemark\footnotetext{The opposite case of either
$\nu_m$ or $\nu_a$ being larger than the observing frequency leads to
rising light curves.}.

Using the allowed ranges of $F_0$, $\tau_0$, and $\nu_0$ we determine
the set of relevant physical parameters: $n_e$, $\gamma_m$, and $B$,
where $B$ is the magnetic field strength.  The radius of the
blastwave, $r$, remains unconstrained (e.g., \citealt{fwk00,bkf04}):
\begin{equation}
B=11.7(p+2)^{-2}F^{-2}_{0,-52}(r_{17}/d_{L,28})^4 \,\,\,{\rm G},
\end{equation}
\begin{equation}
\gamma_m=6.7(p+2)F_{0,-52}\nu^{1/2}_{0,9}(r_{17}/d_{L,28})^{-2},
\end{equation}
\begin{equation}
n_e=3.6\times10^{10}c_n\eta_1F^3_{0,-52}\nu^{(1-p)/2}_{0,9}
\tau_{0,32}r^{-1}_{17}(r_{17}/d_{L,28})^{-6} \,\,\,{\rm cm}^{-3},
\end{equation}
\begin{equation}
c_n=(1.67\times10^3)^{-p}(5.4\times 10^2)^{(1-p)/2}(p+2)^2 /(p-1)
\end{equation}
where $d_{L,28}$ is the luminosity distance in units of $10^{28}$ cm
assuming the standard cosmological parameters ($H_0=71$ km s$^{-1}$
Mpc$^{-1}$, $\Omega_M=0.27$, and $\Omega_{\Lambda}=0.73$), and $\eta$
is the reciprocal of the thickness of the emitting shell.  

The unknown radius of the blastwave can be constrained by introducing
a relationship between $E_{\rm ST}=n_e m_p(r/1.05)^5[t_{NR}/
(1+z)]^{-2}$ and the energy in the electrons and the magnetic field.
We use the condition that at most half of the blastwave energy is
available for accelerating electrons and producing the magnetic field,
i.e., $(E_B + E_e)\lesssim E_{\rm ST}/2$.  The total energy in the
accelerated electrons is $E_e=[(p-1)/(p-2)]n_e\gamma_mc^2V$, while the
energy in the magnetic field is $E_B=B^2V/8\pi$, where $V=4\pi
r^2/\eta$ is the volume of the synchrotron emitting shell.  The energy
budget is minimized near equipartition (i.e., $E_e\approx E_B$), and
we use this constraint to determine the minimum required energy (e.g.,
\citealt{fwk00,bkf04}); this conclusion was verified with radio
interferometric measurements of the size of GRB\,030329
\citep{tfb+04,fsk+05}.

Finally, using the inferred radius for each possible solution, we
require for self-consistency that $\beta\lesssim 1$, where
$\beta=2r(1+z)/5ct$.  The resulting $\beta$ distributions are shown in
Figure~\ref{fig:beta}.  These results indicate that most of the
detected bursts obey the self-consistency requirement, although we
reject GRBs 000926, 020819, and 021004 from the analysis since
$\gtrsim 50\%$ of their allowed solutions lead to relativistic
velocities at the time of the observations.  This does not rule out
that the Sedov-Taylor solution is applicable, but simply indicates
that additional observations are required to narrow down the range of
allowed solutions.  We furthermore find that the bulk of the upper
limits do not rule out relativistic expansion, and we therefore do not
use these limits in the energy distribution analysis below.

\section{The Distribution of GRB Kinetic Energies}
\label{sec:res}

\begin{figure}
\includegraphics[width=3.2in]{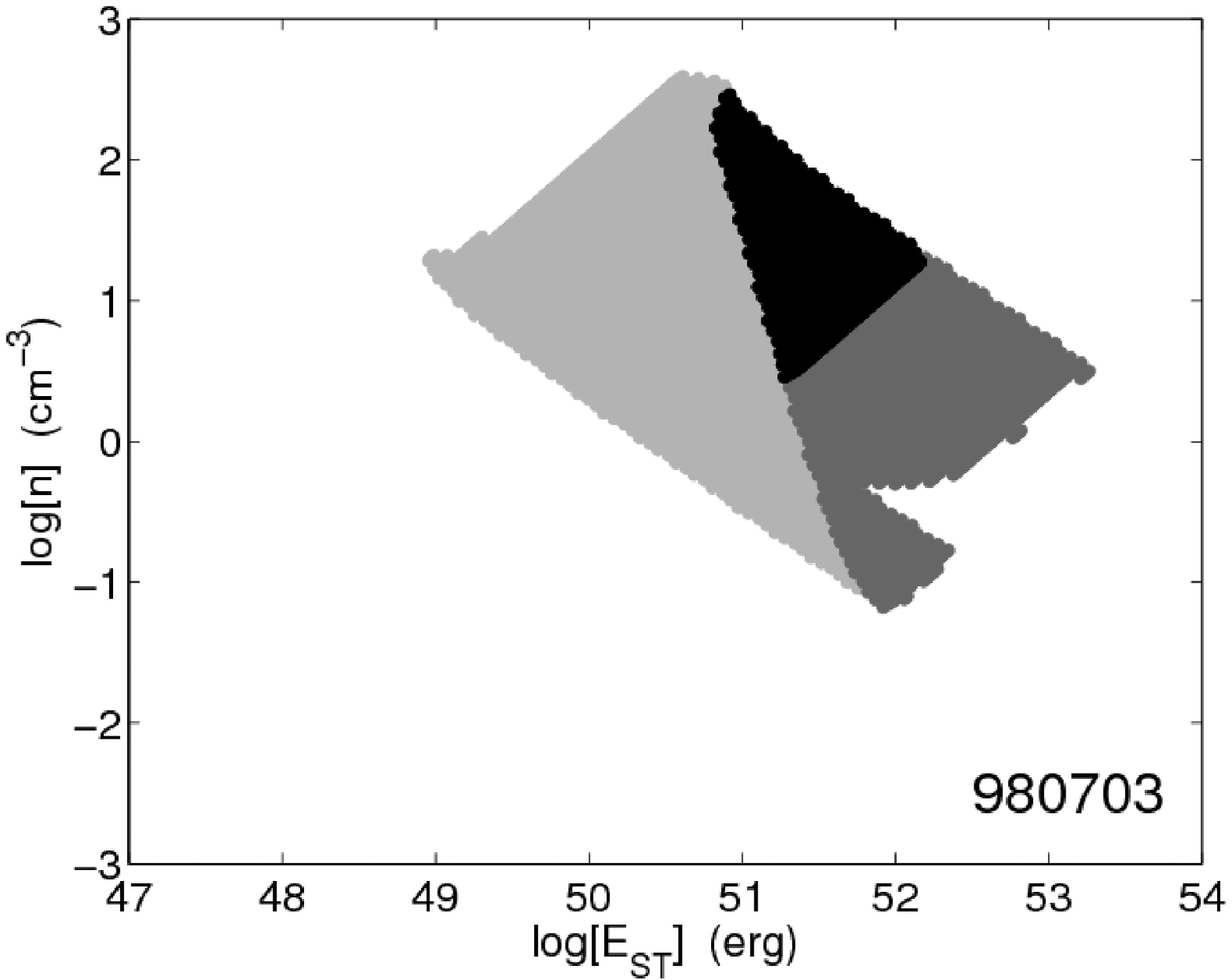}
\includegraphics[width=3.2in]{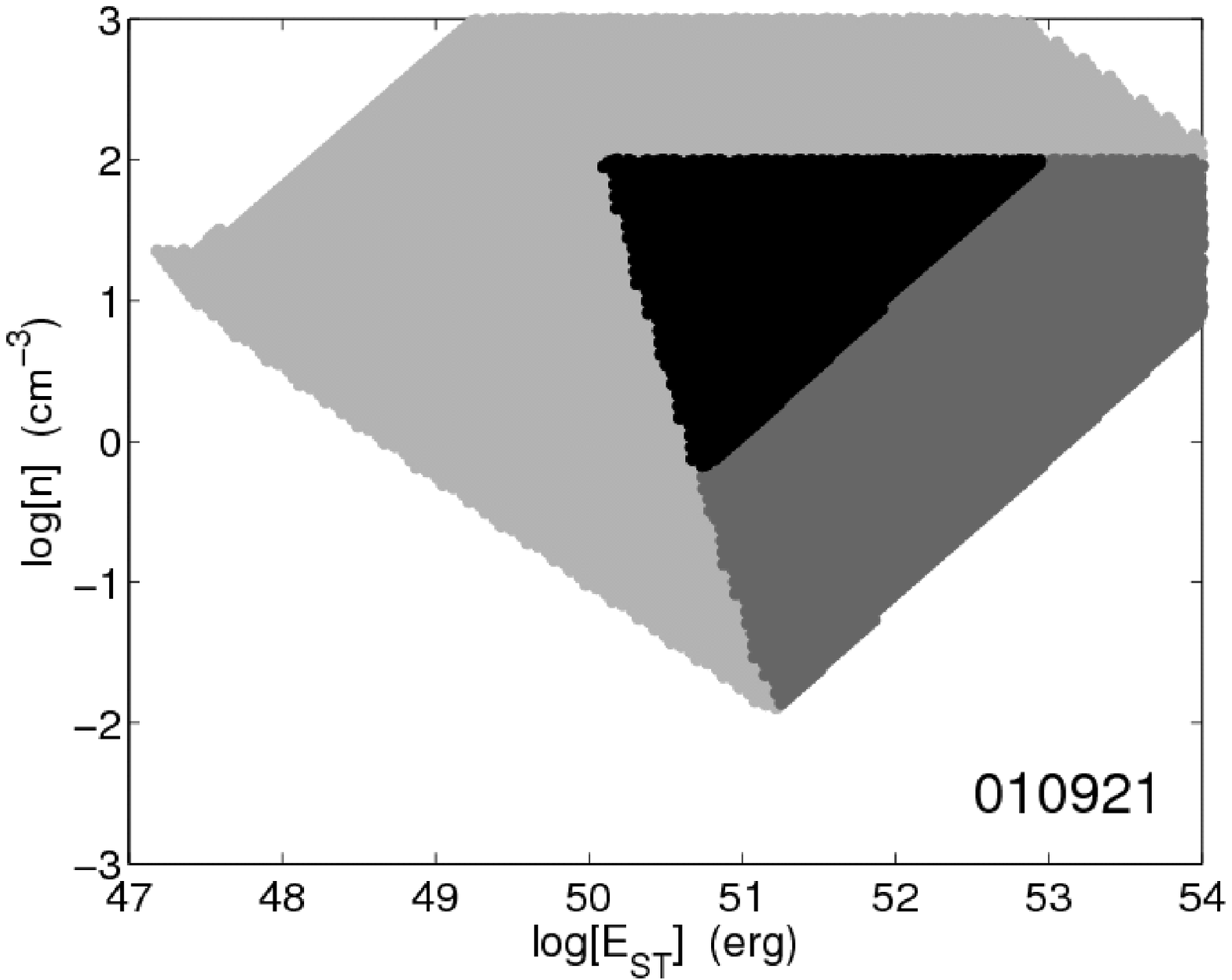} \\
\caption{Electron number density plotted against kinetic energy for
two representative cases.  The light gray regions indicate the
phase-space that leads to a predicted flux density in agreement with
the observed values.  The medium gray regions encompass the subset of
solutions that satisfy the condition $(E_B + E_e)\lesssim E_{\rm
ST}/2$.  The black regions marks the subset of solutions that satisfy
$\beta<1$ in the Sedov-Taylor framework.  {\it Left:} Group A burst
with a well-defined spectral peak.  {\it Right:} Group B burst with a
single frequency detection for which we use the additional limit that
$n<100$ cm$^{-3}$.  This figure highlights the significant advantage
of measuring the spectral peak.}
\label{fig:examples} 
\end{figure}

\begin{figure}
\includegraphics[width=1.55in]{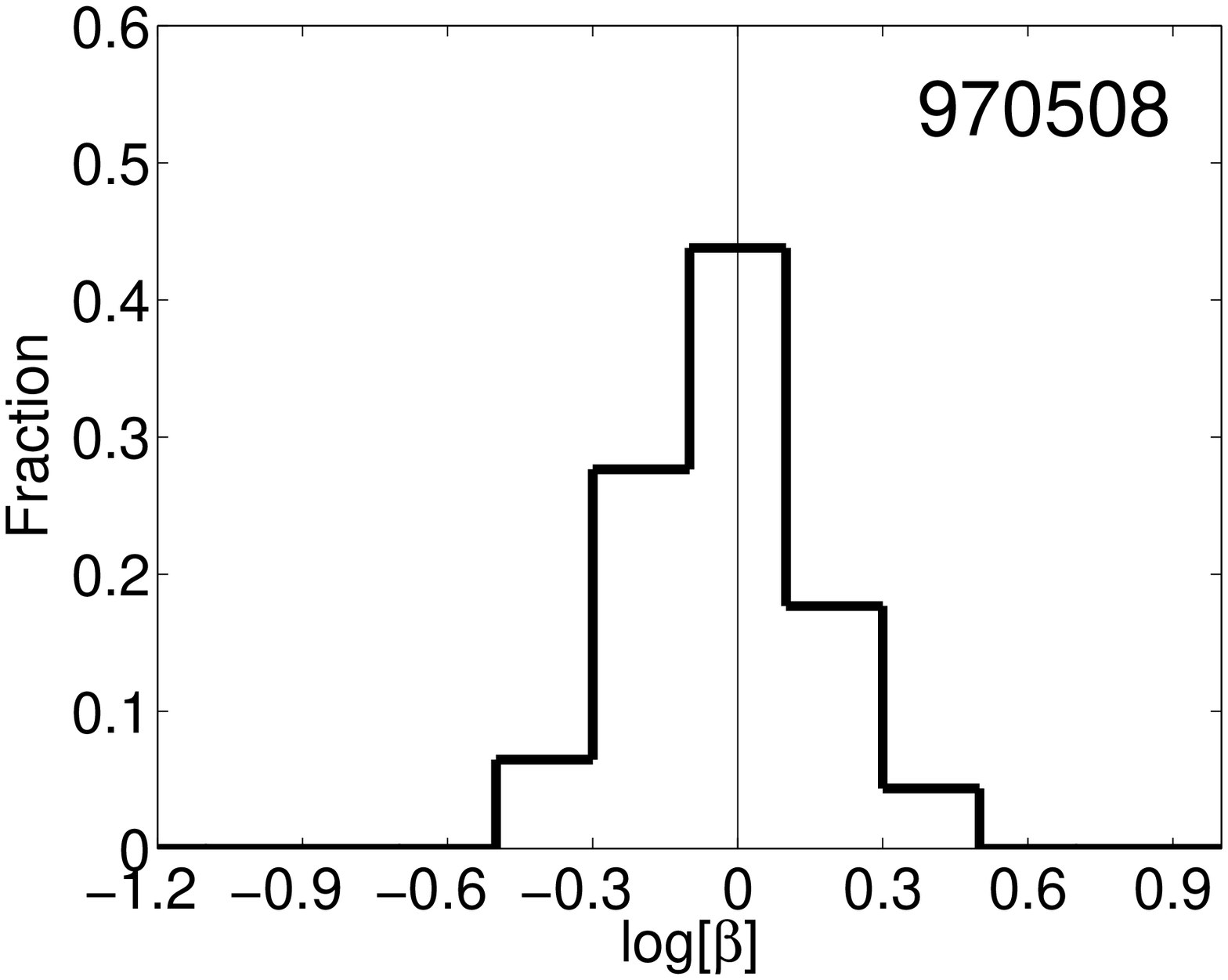} 
\includegraphics[width=1.55in]{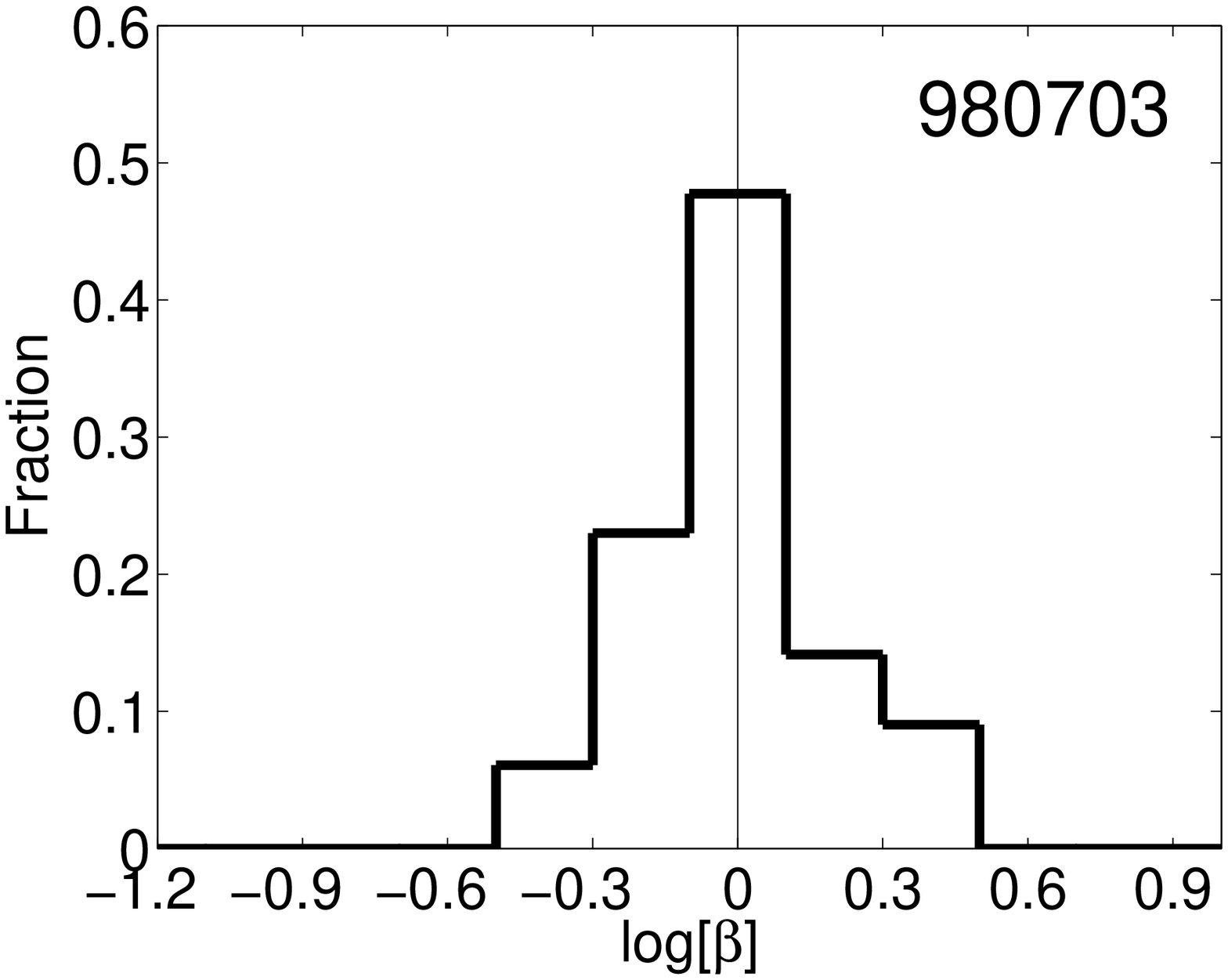} 
\includegraphics[width=1.55in]{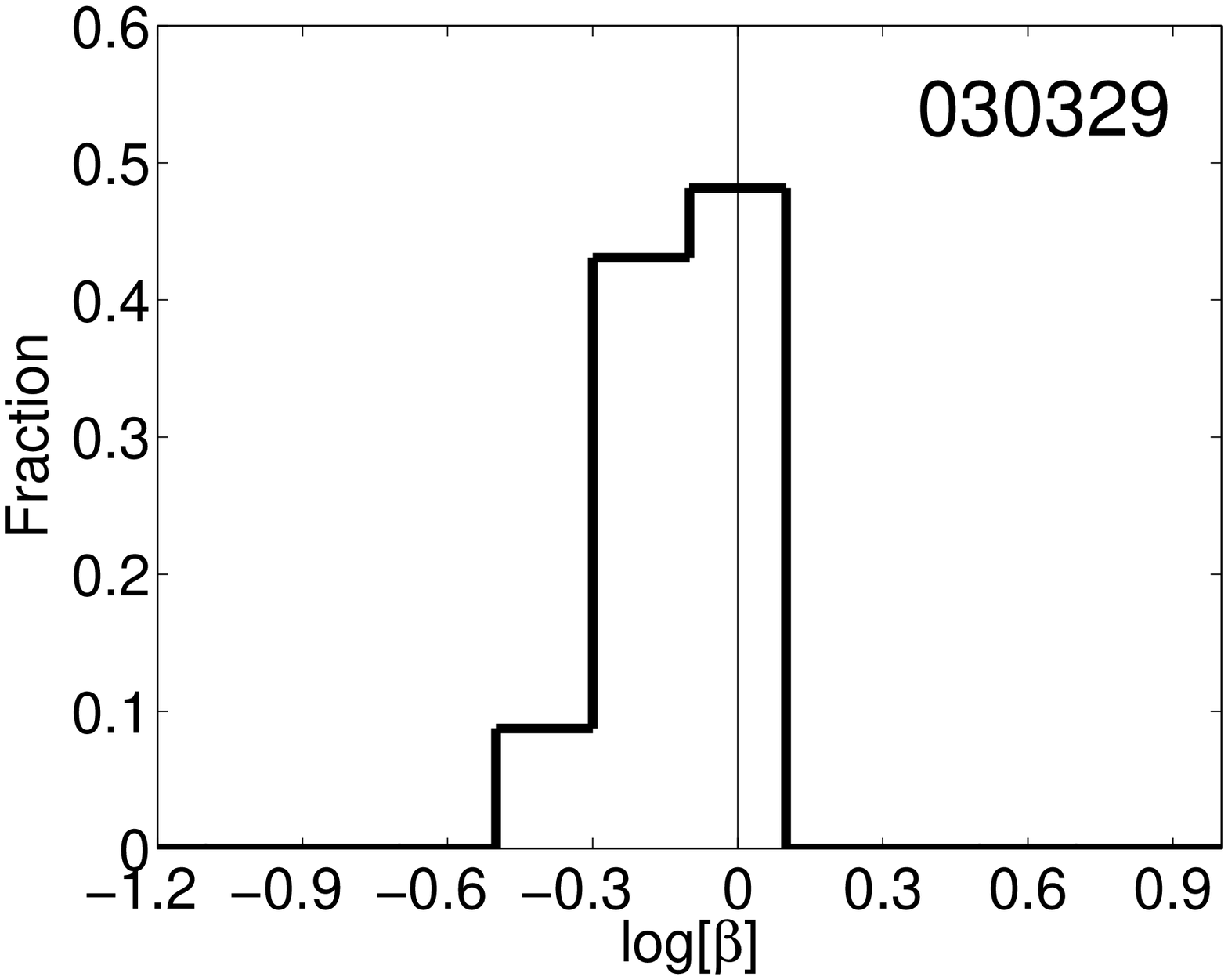} 
\includegraphics[width=1.55in]{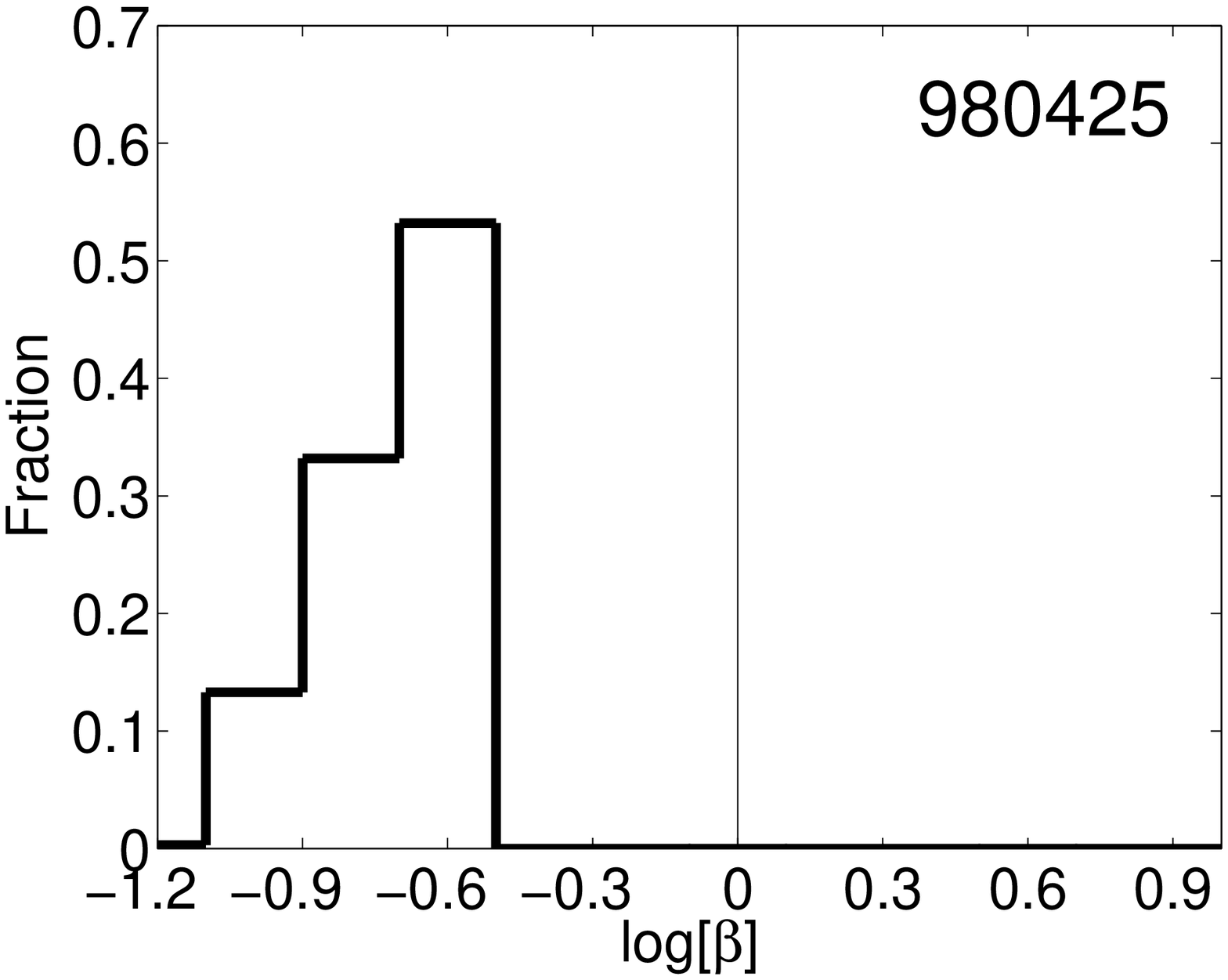} \\
\includegraphics[width=1.55in]{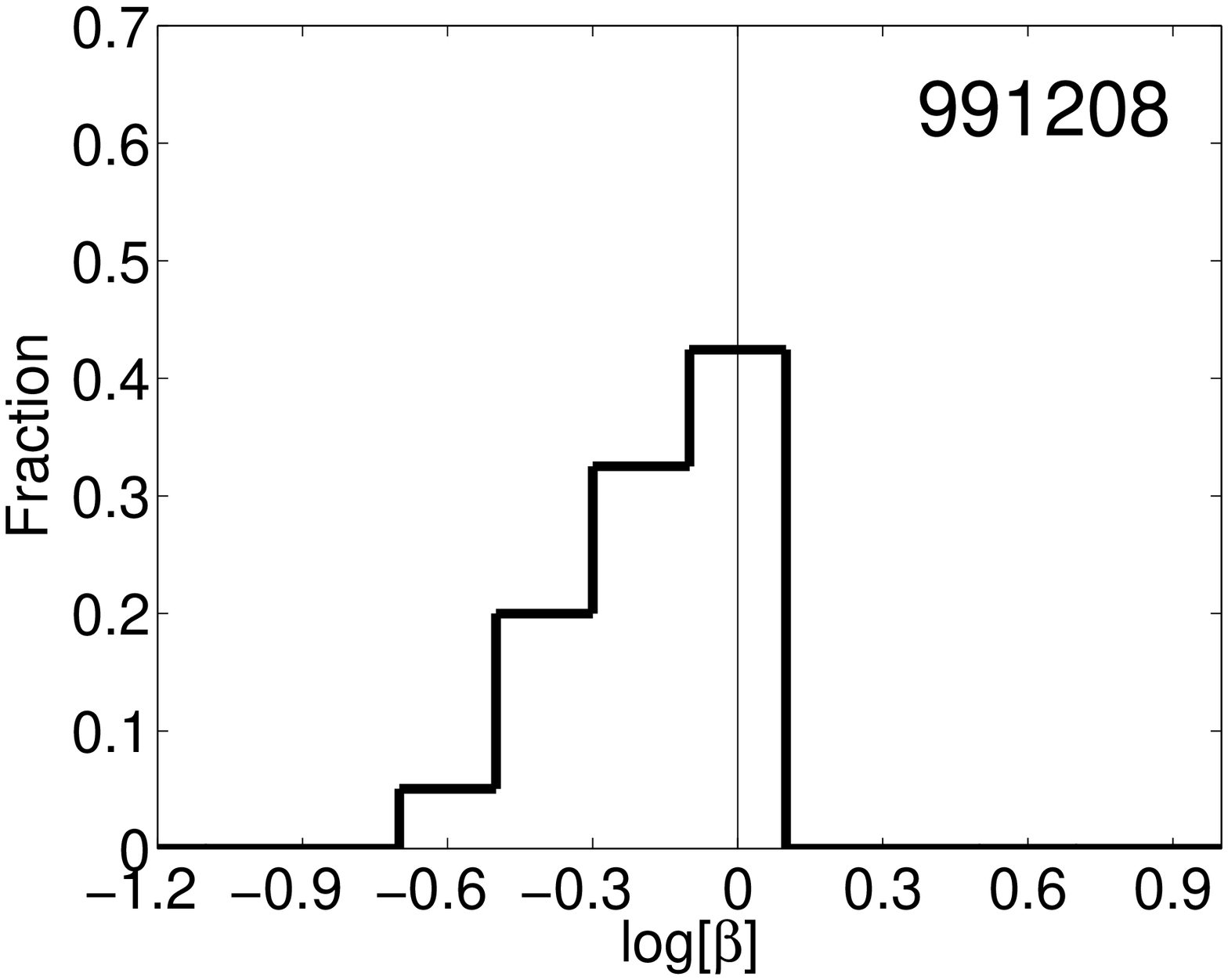} 
\includegraphics[width=1.55in]{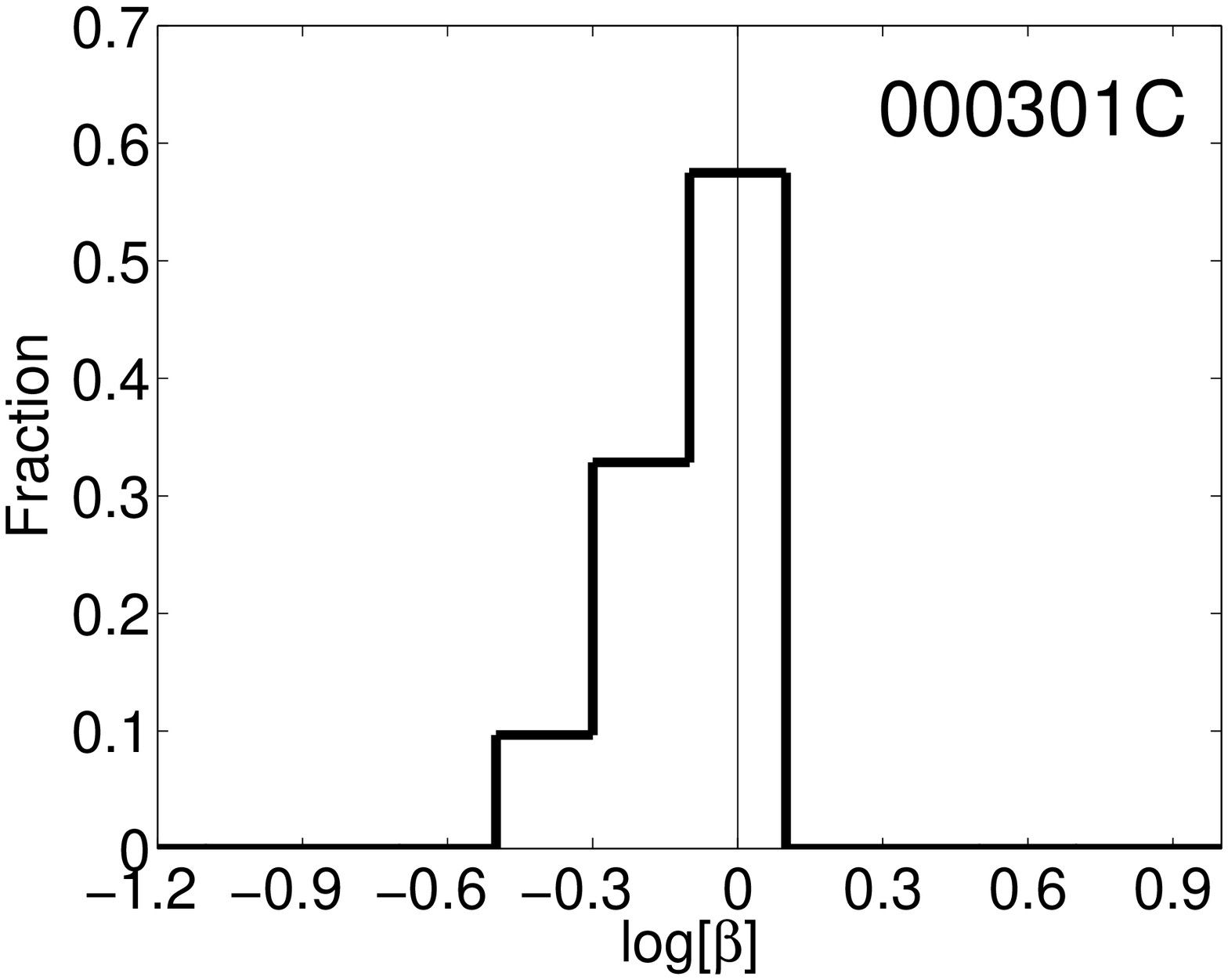} 
\includegraphics[width=1.55in]{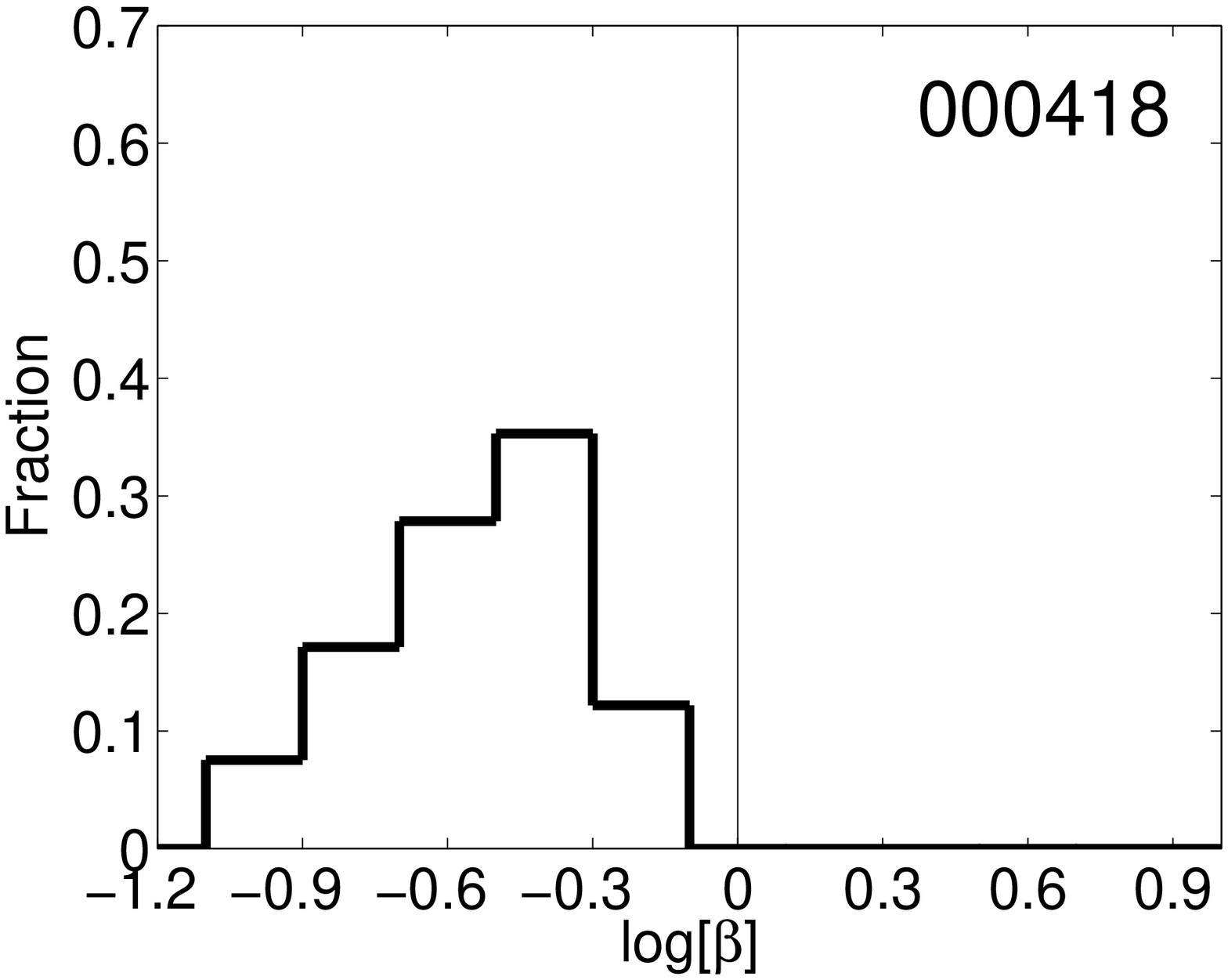} 
\includegraphics[width=1.55in]{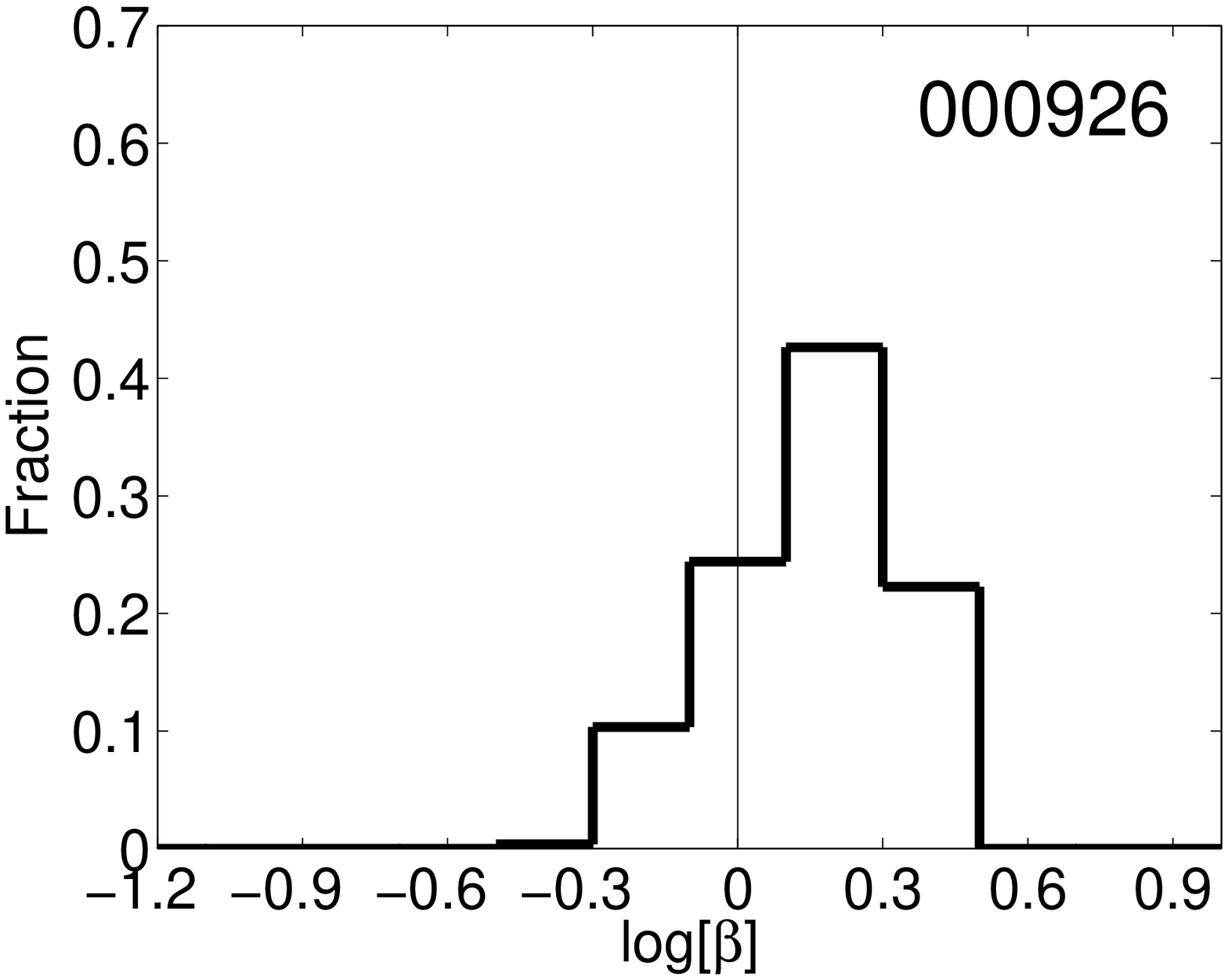} \\
\includegraphics[width=1.55in]{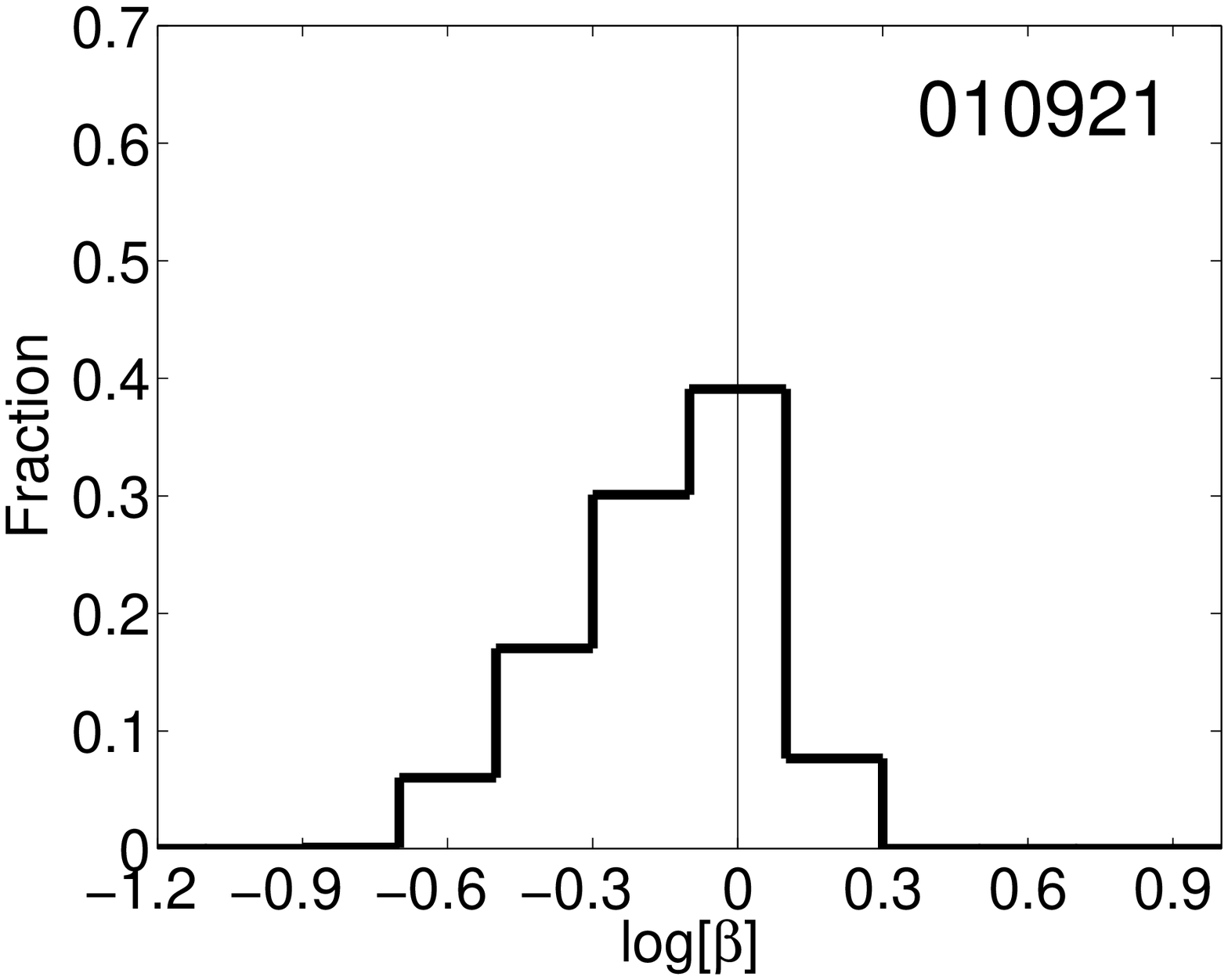} 
\includegraphics[width=1.55in]{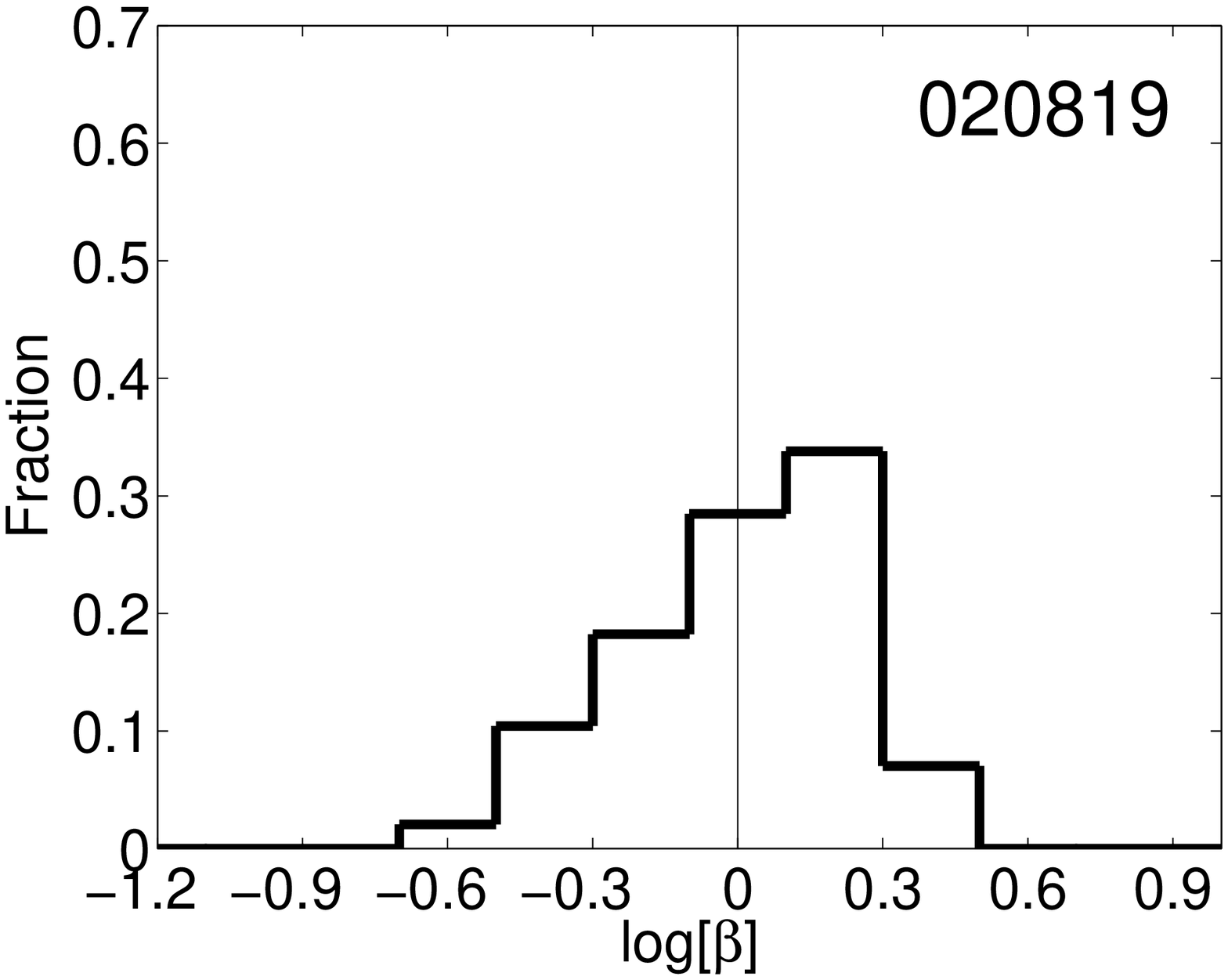} 
\includegraphics[width=1.55in]{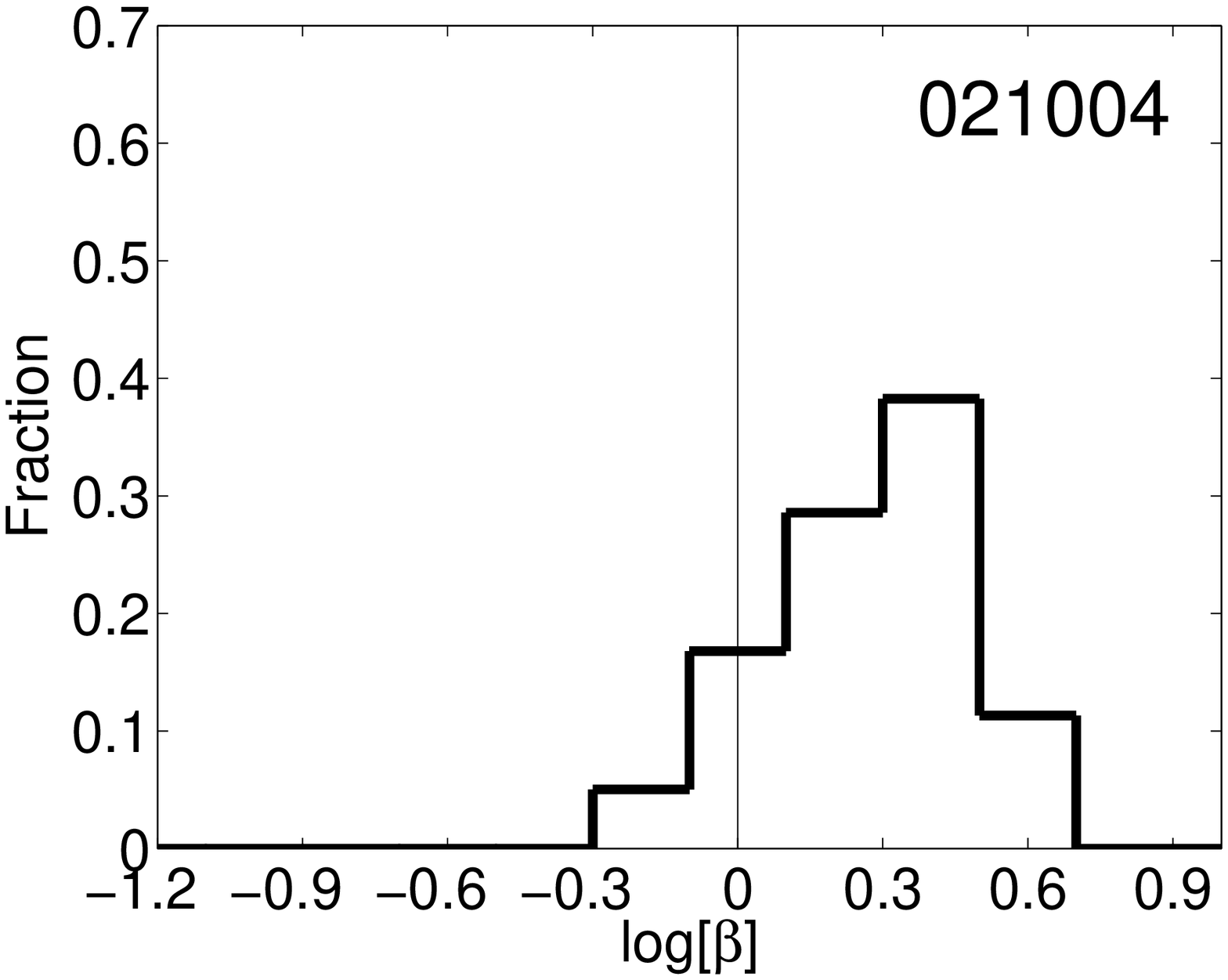} 
\includegraphics[width=1.55in]{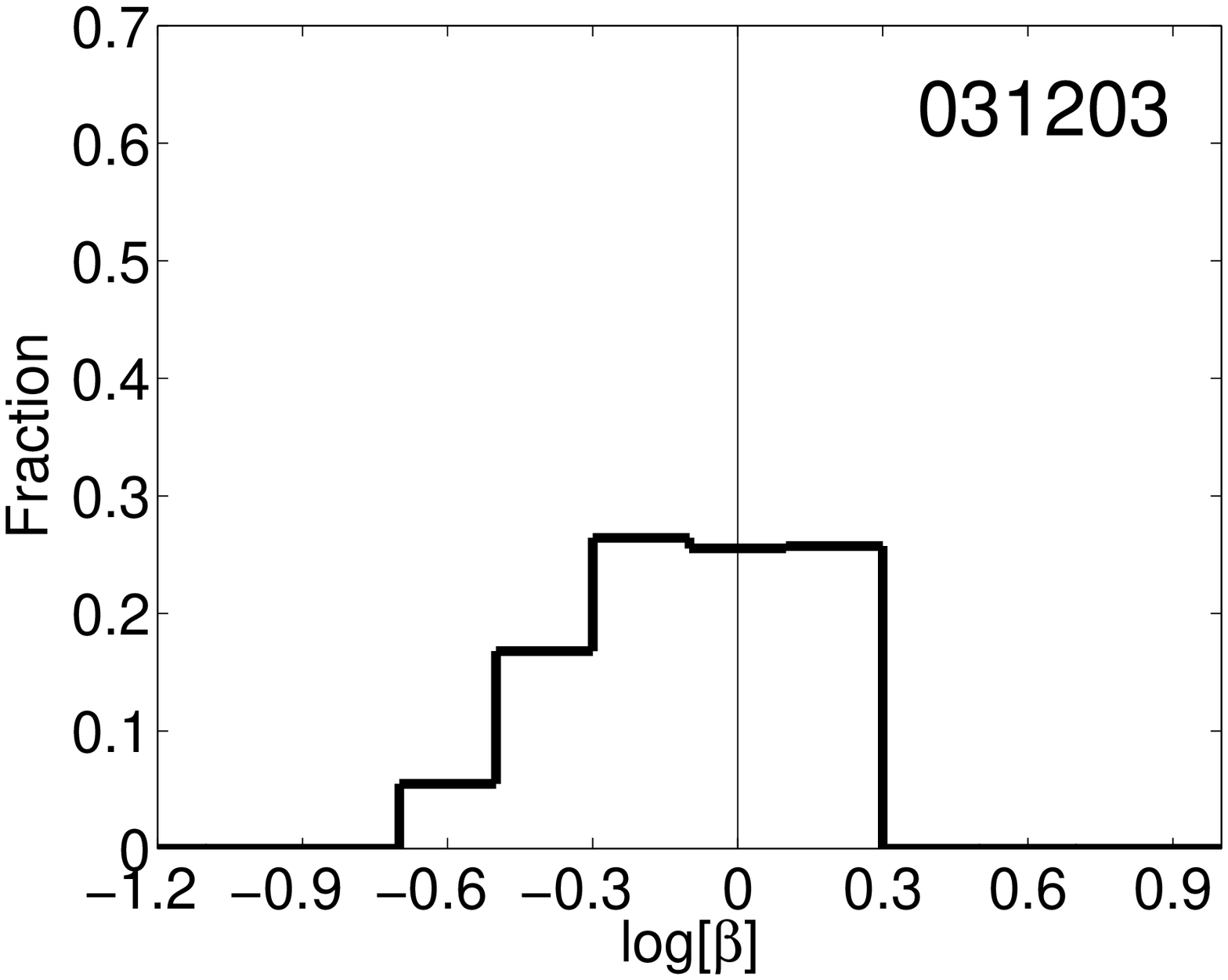} \\
\includegraphics[width=1.55in]{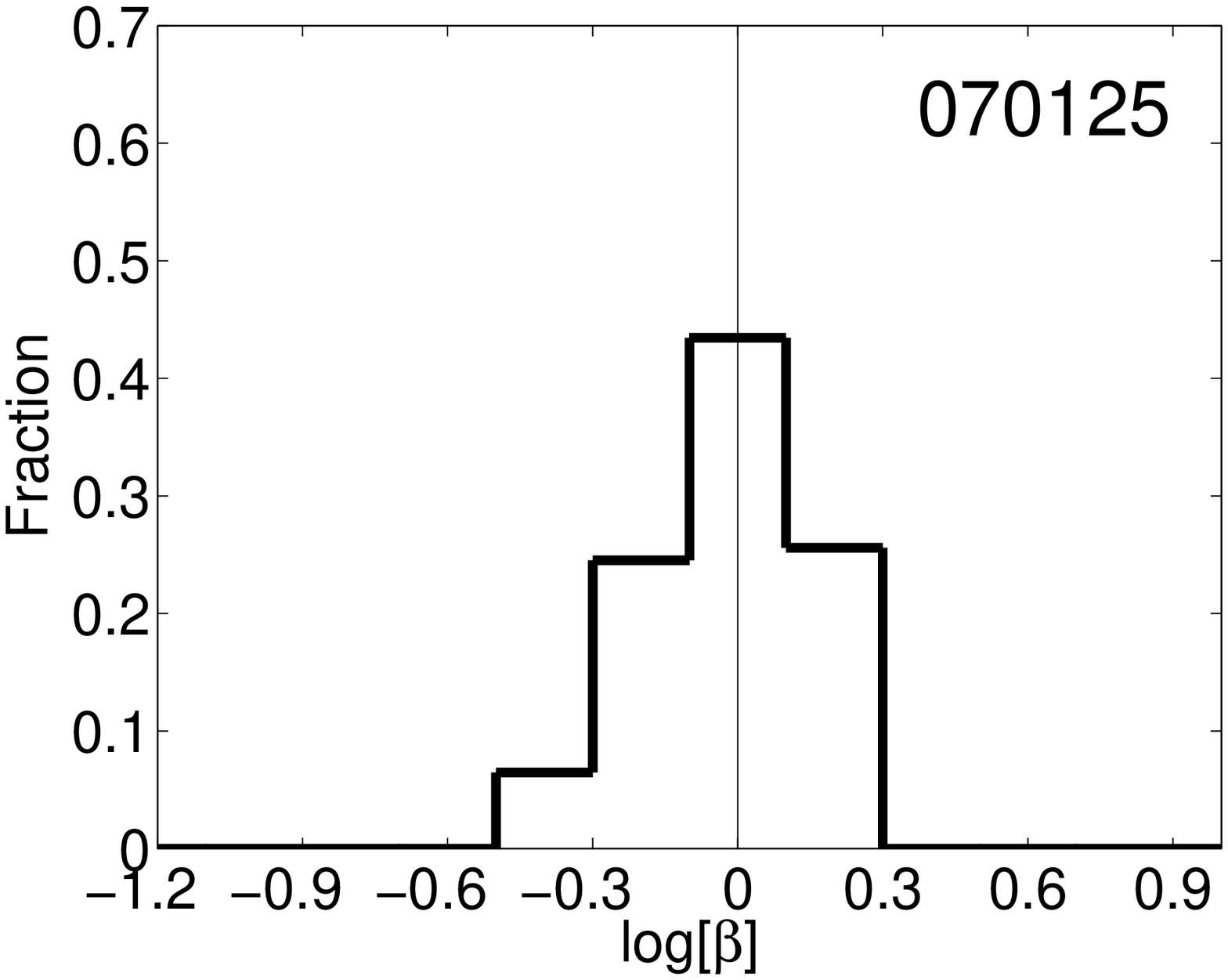} 
\includegraphics[width=1.55in]{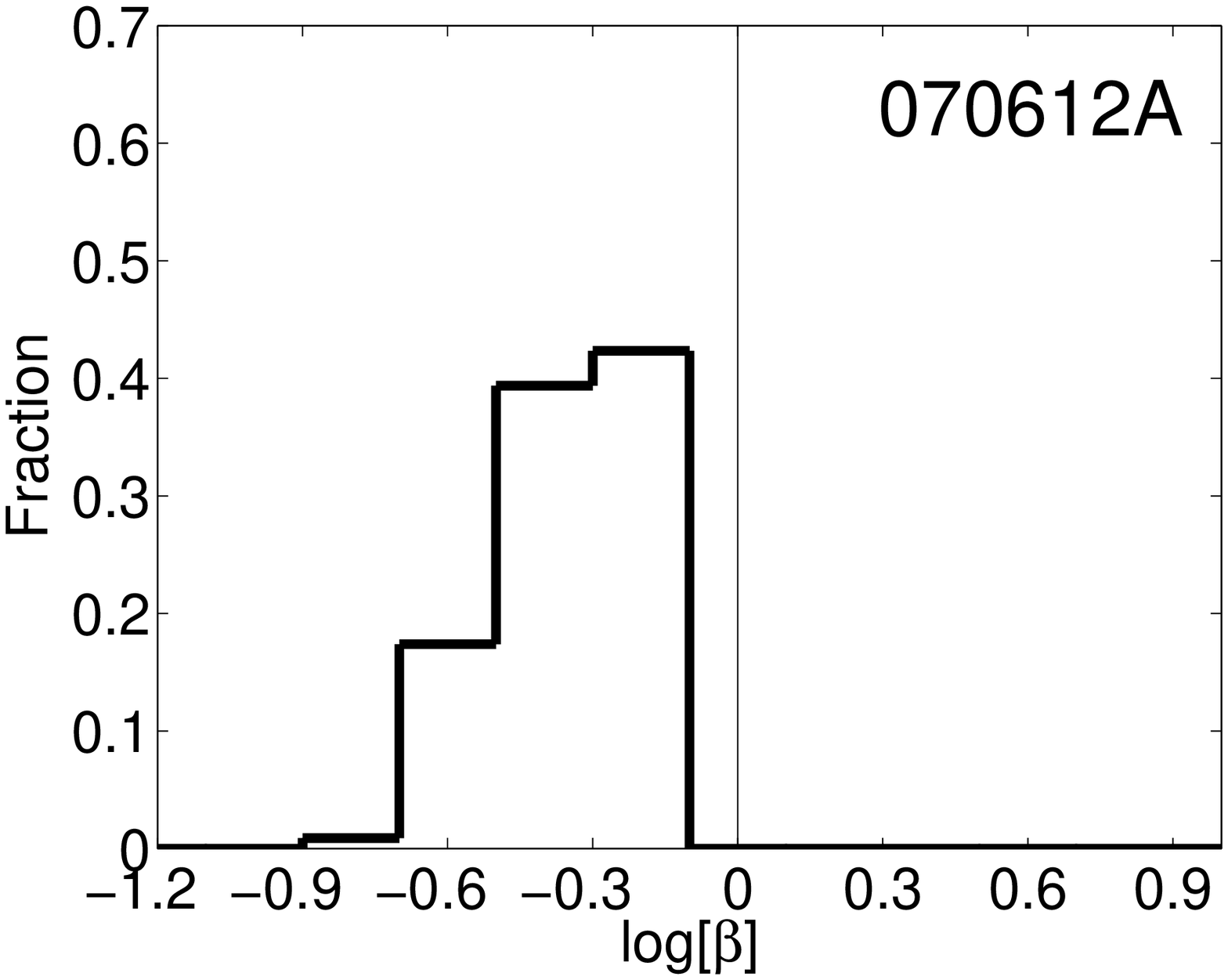} 
\includegraphics[width=1.55in]{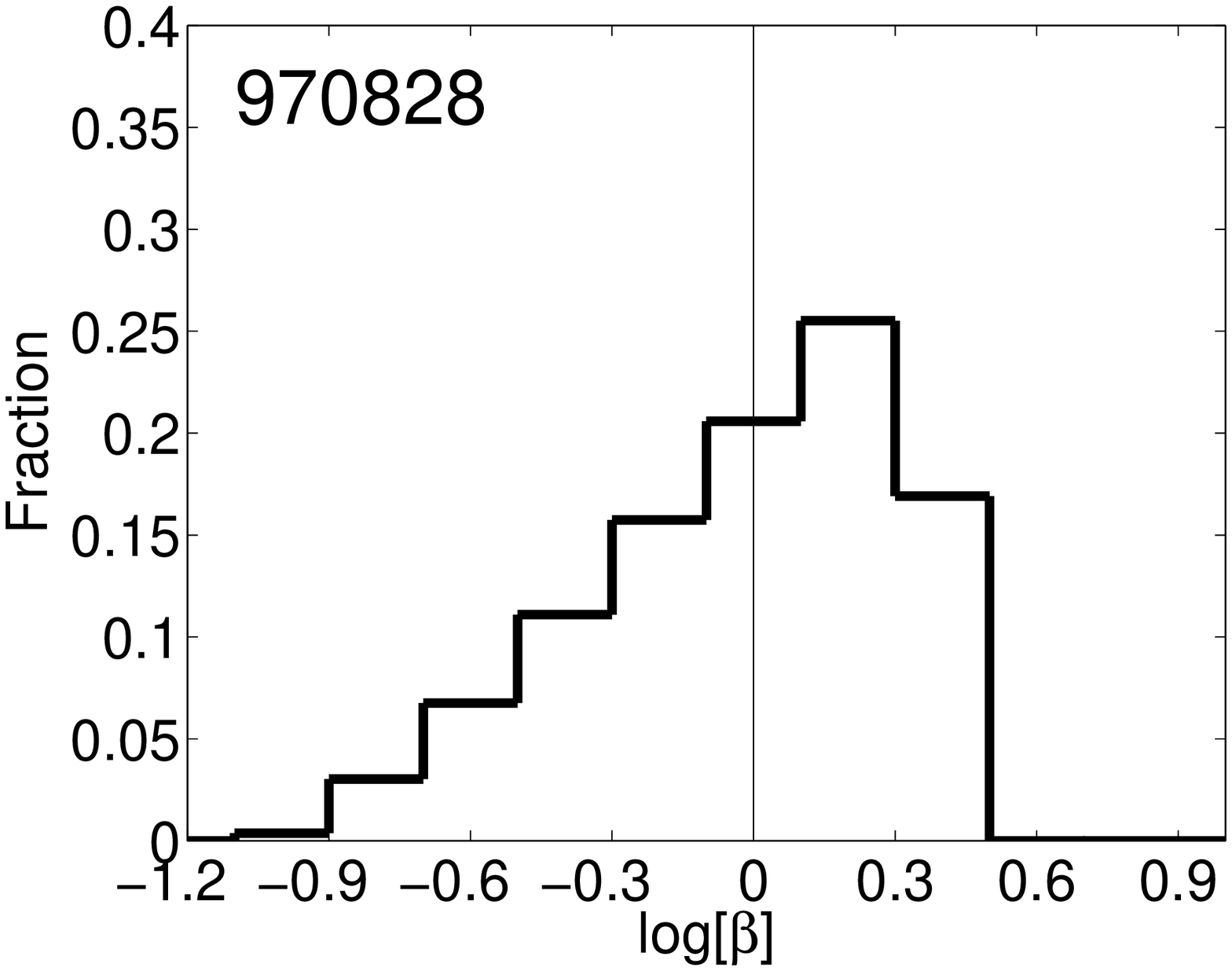} 
\includegraphics[width=1.55in]{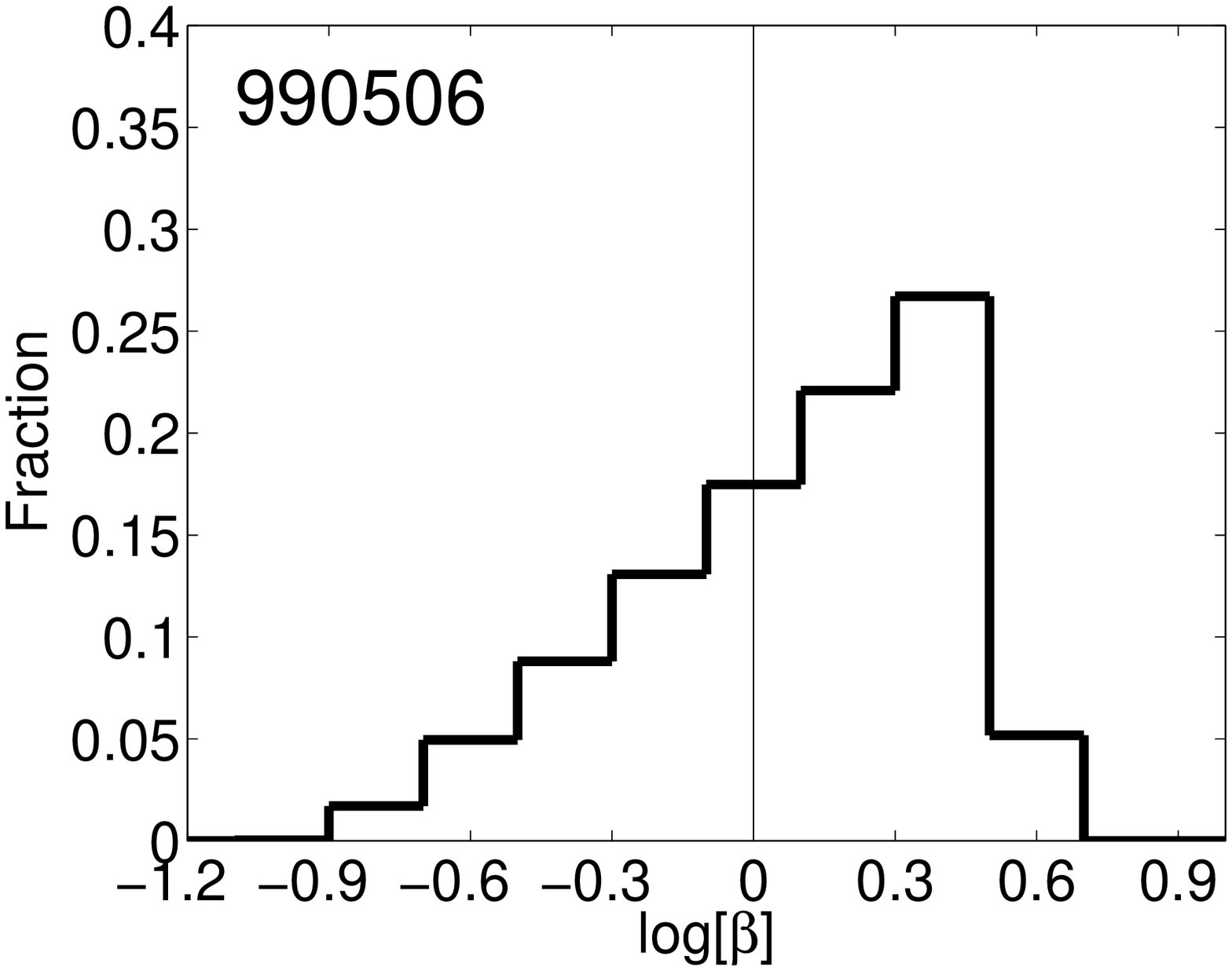} \\
\includegraphics[width=1.55in]{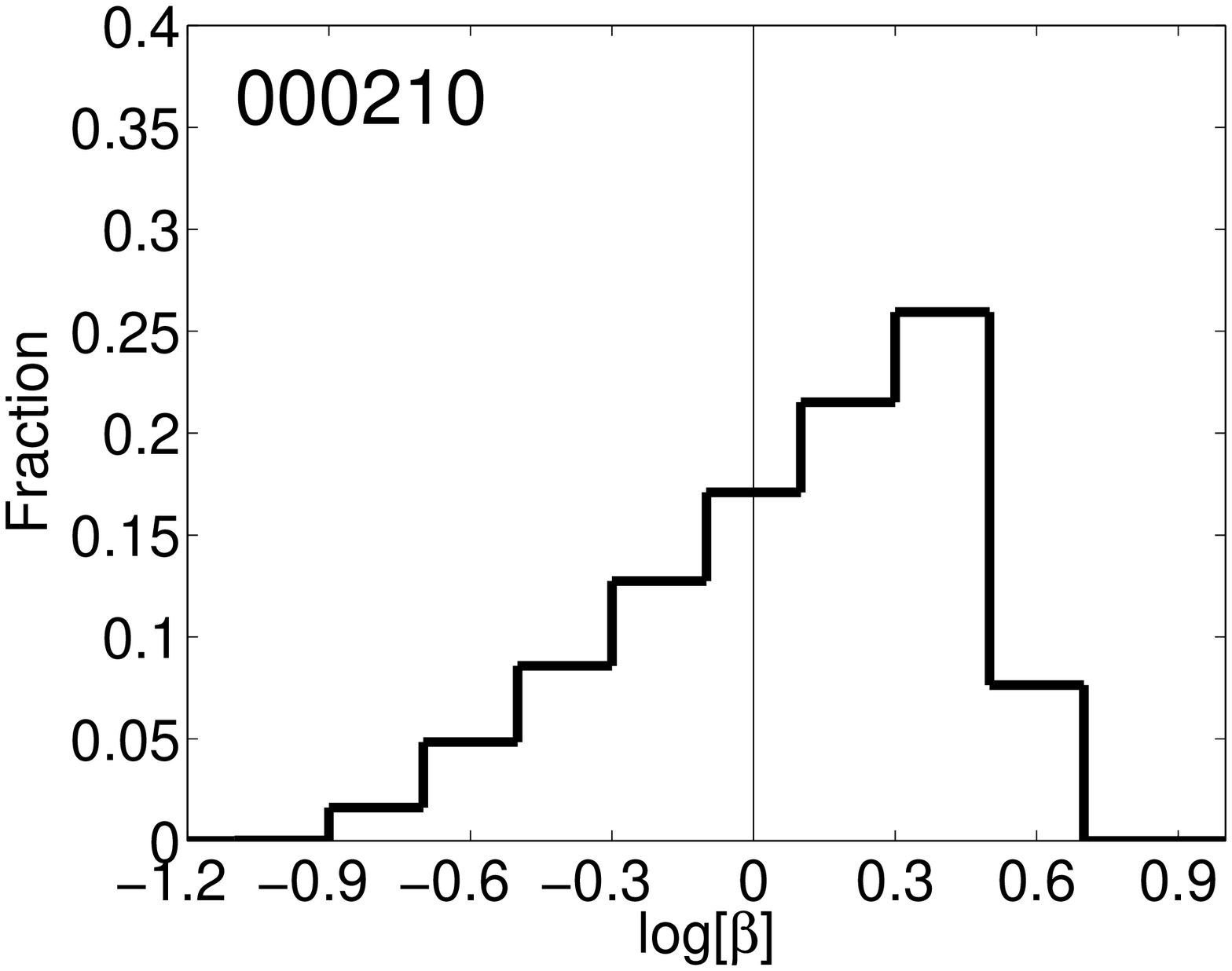} 
\includegraphics[width=1.55in]{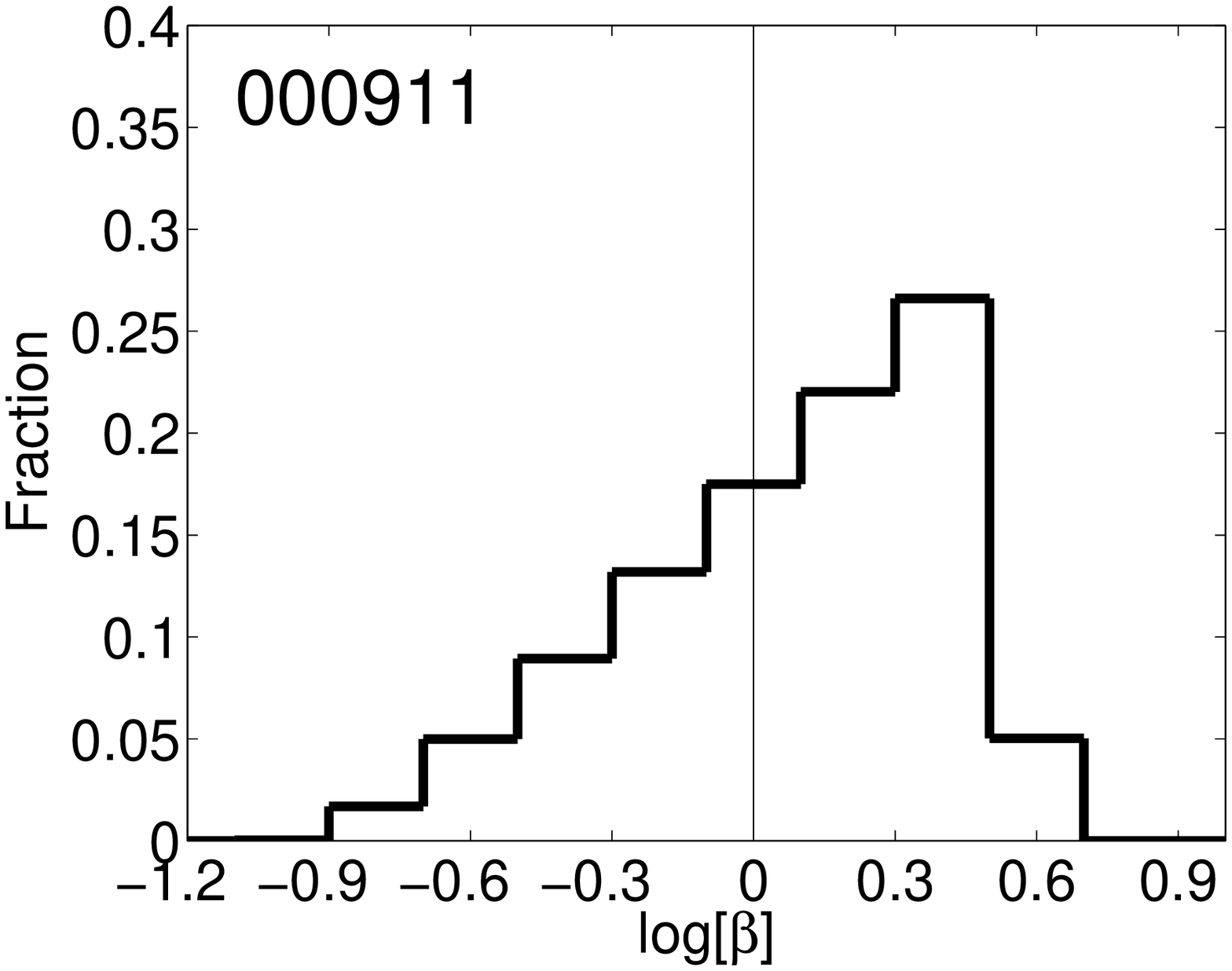} 
\includegraphics[width=1.55in]{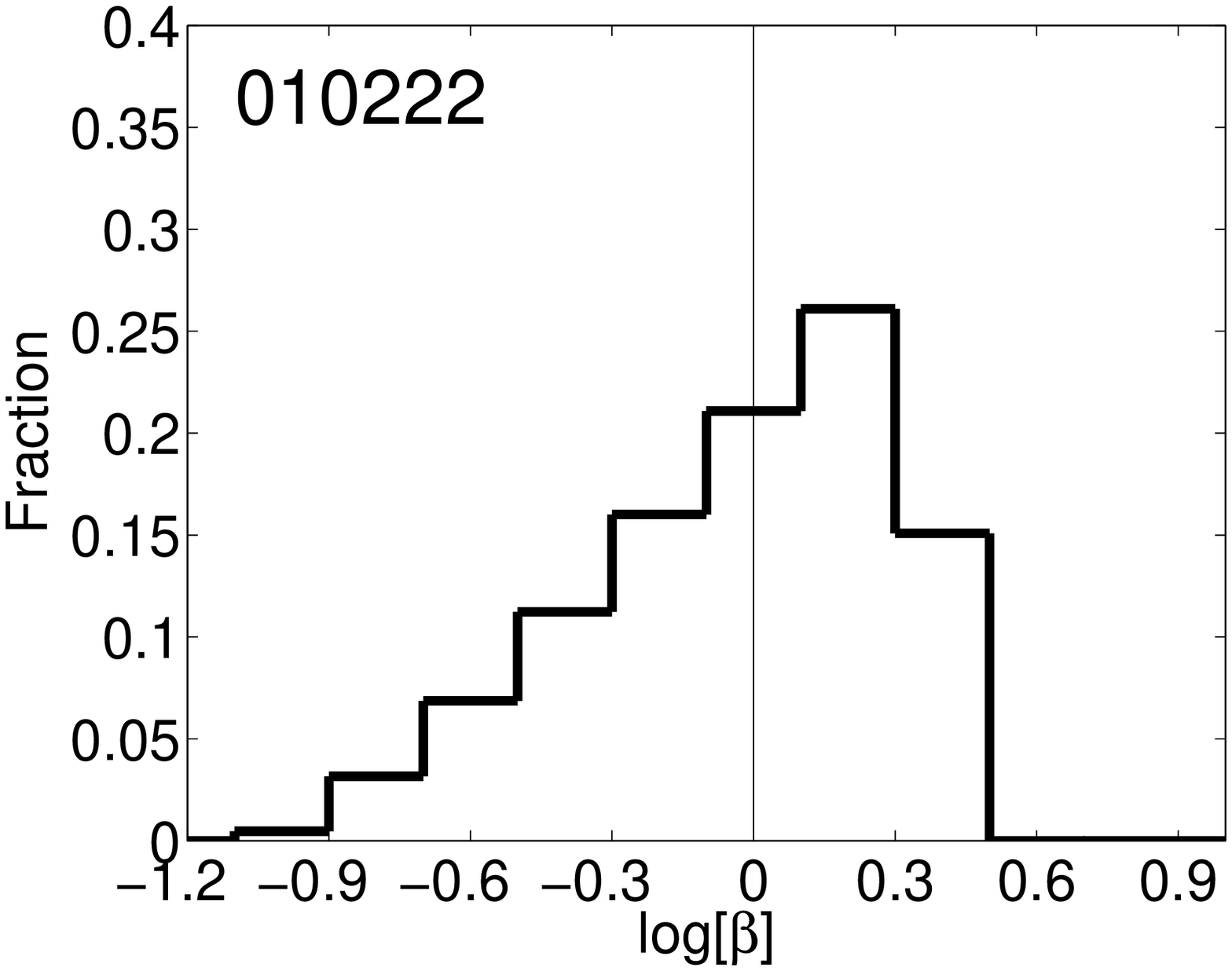} 
\includegraphics[width=1.55in]{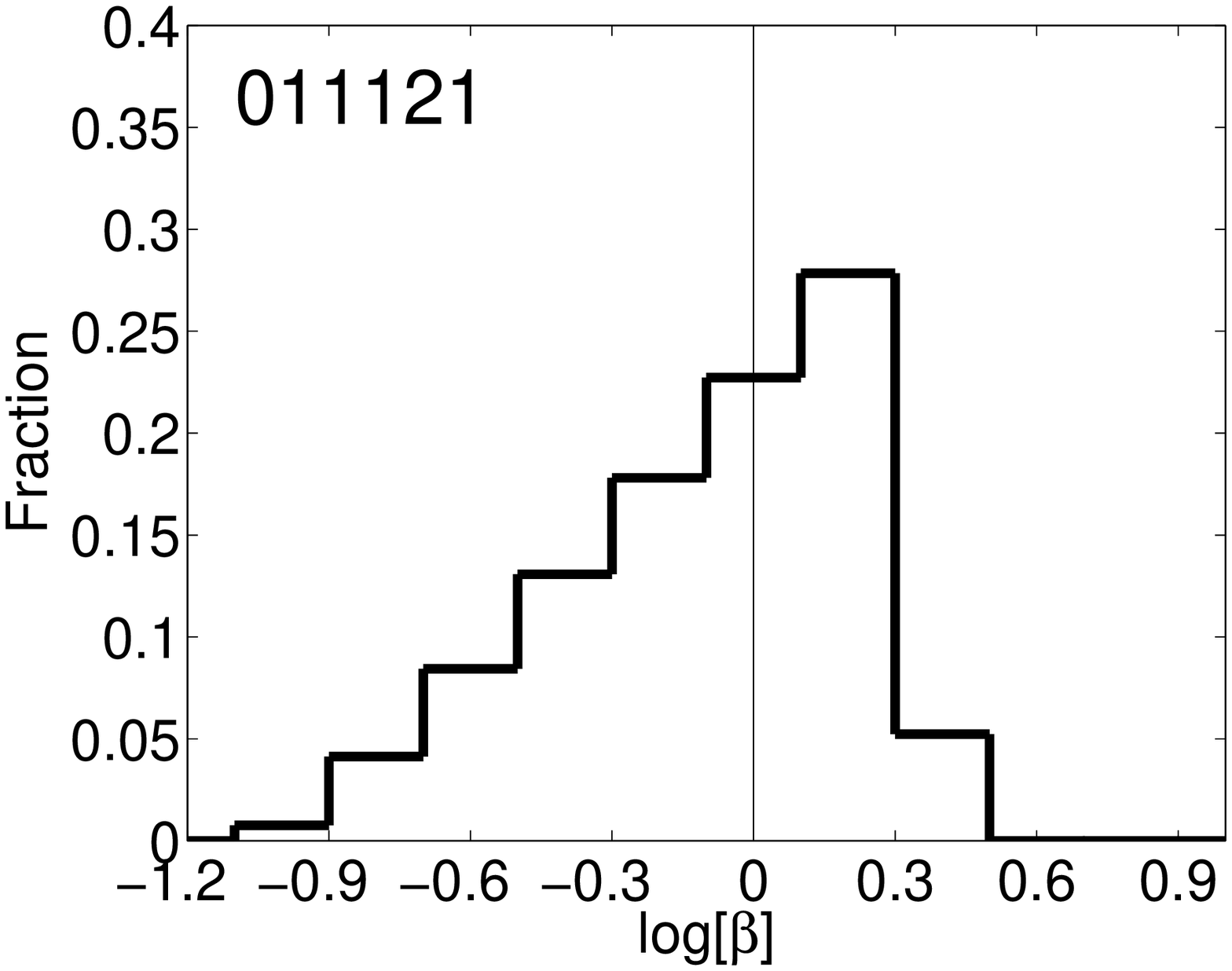} \\
\includegraphics[width=1.55in]{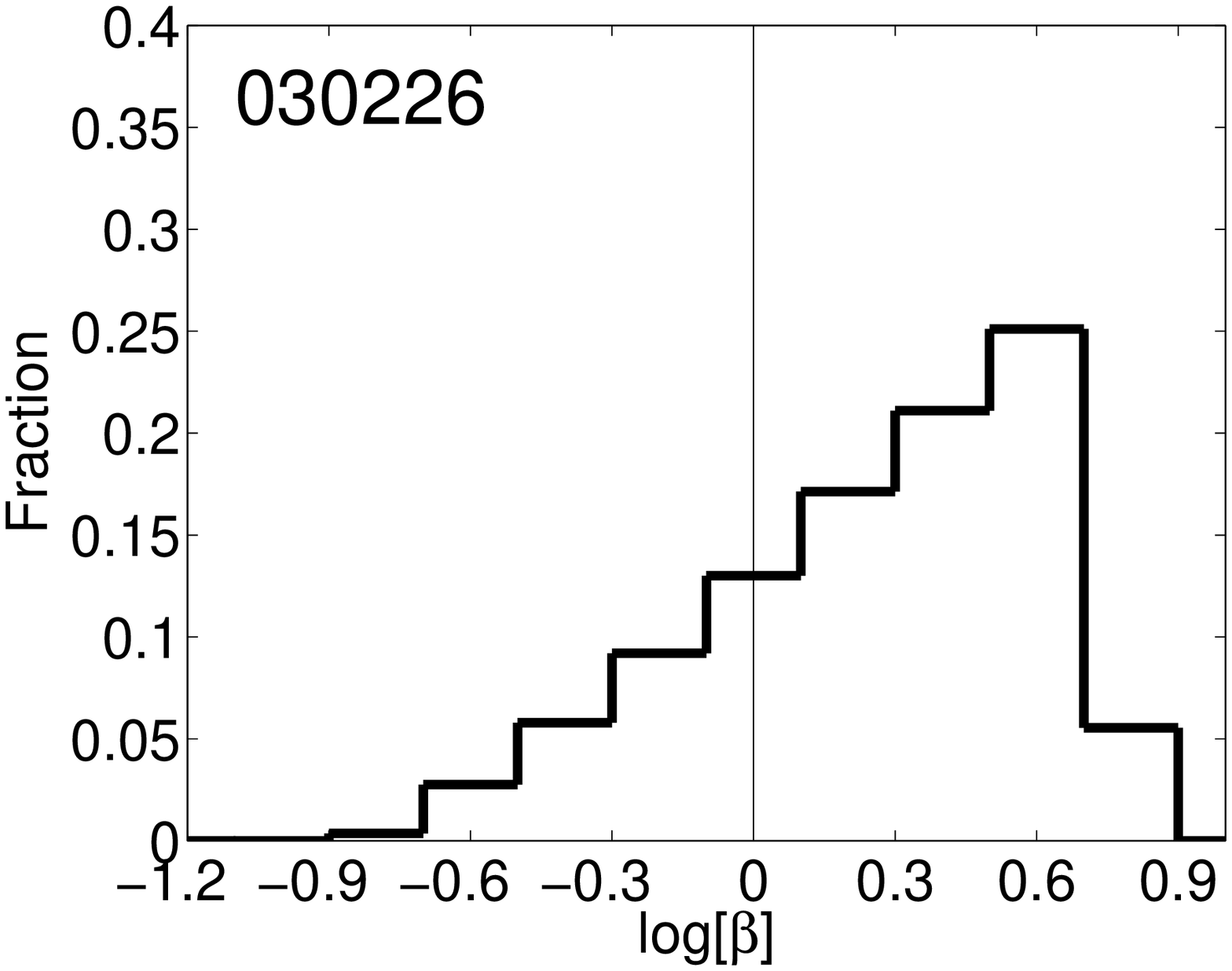} 
\includegraphics[width=1.55in]{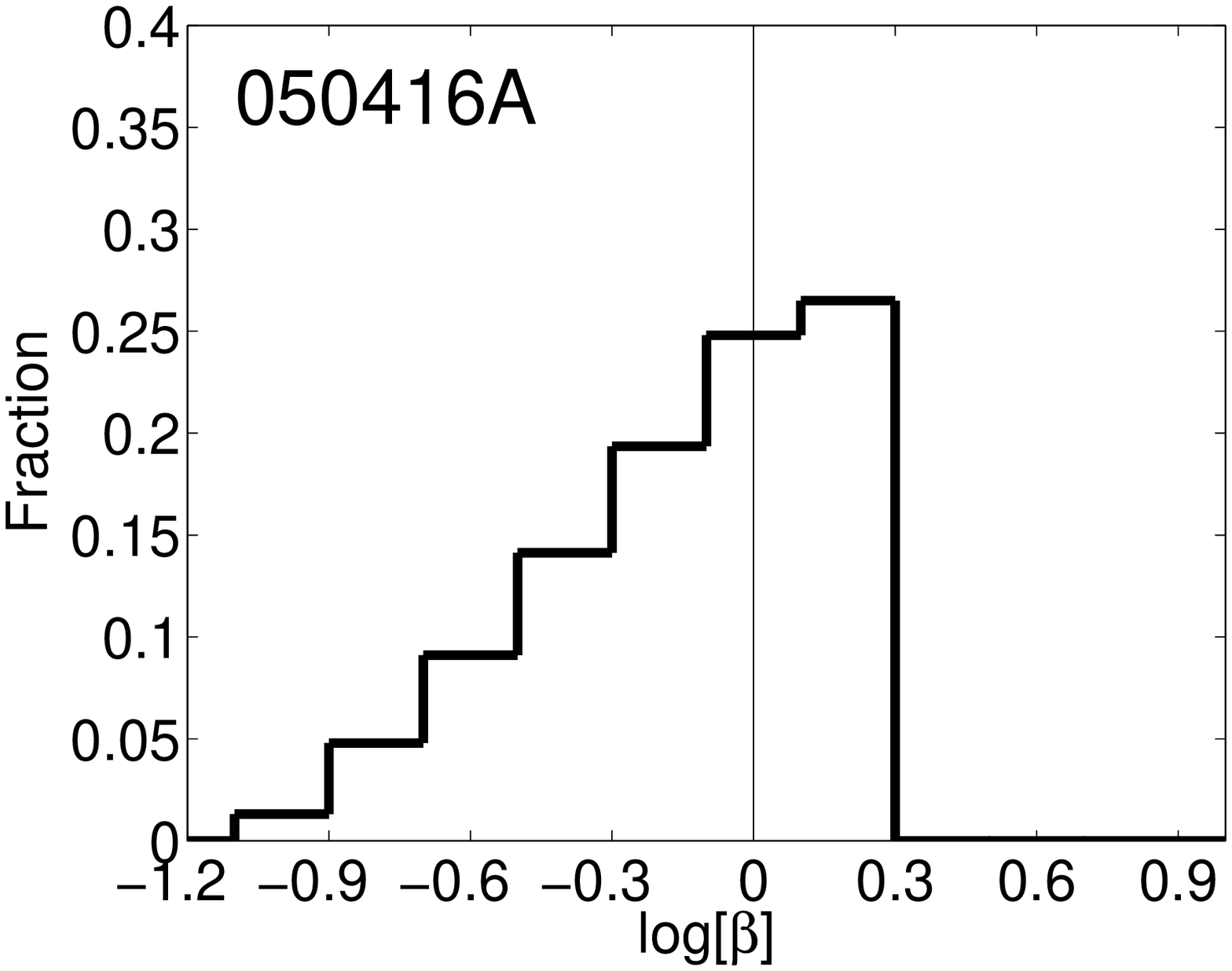} 
\includegraphics[width=1.55in]{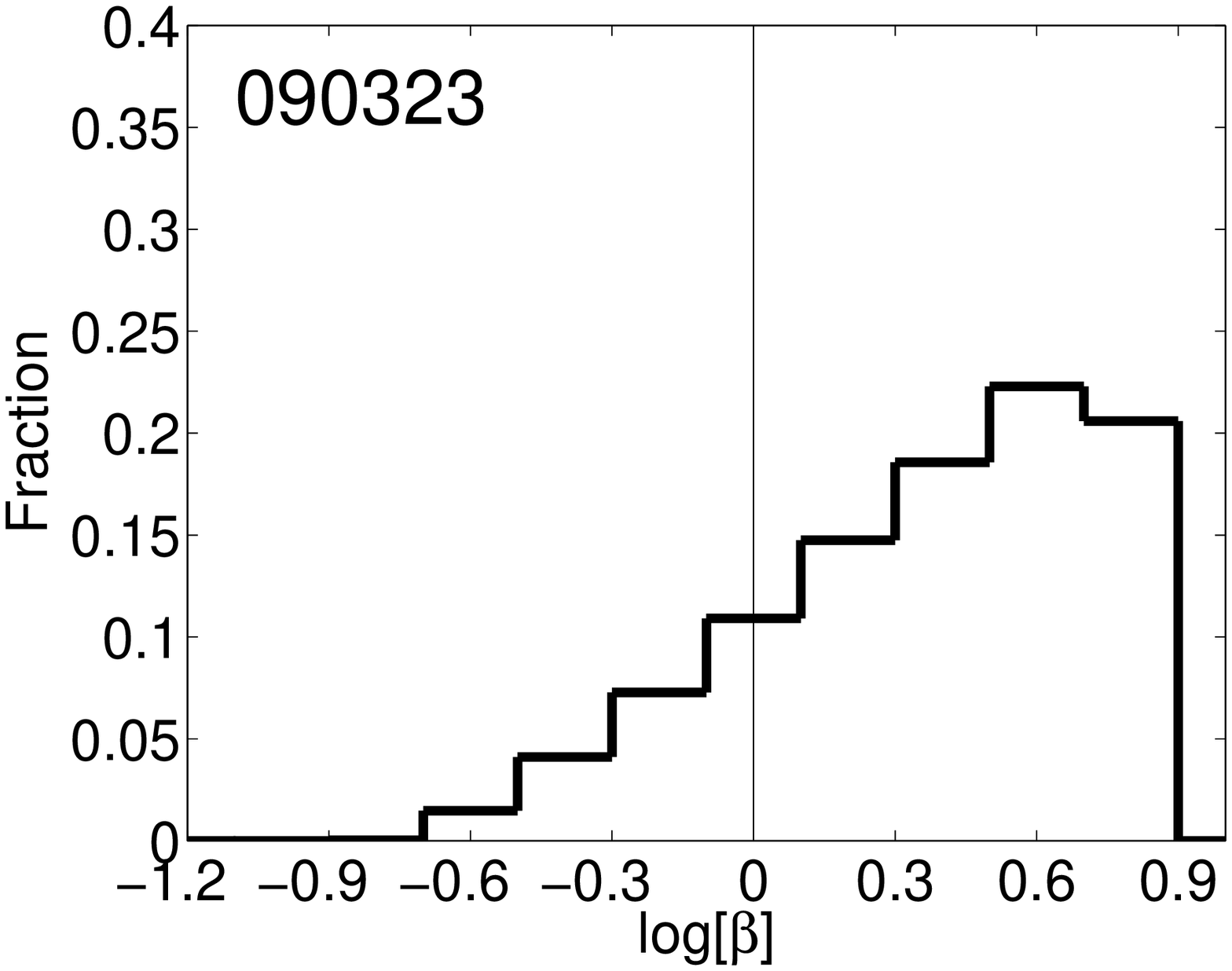} 
\includegraphics[width=1.55in]{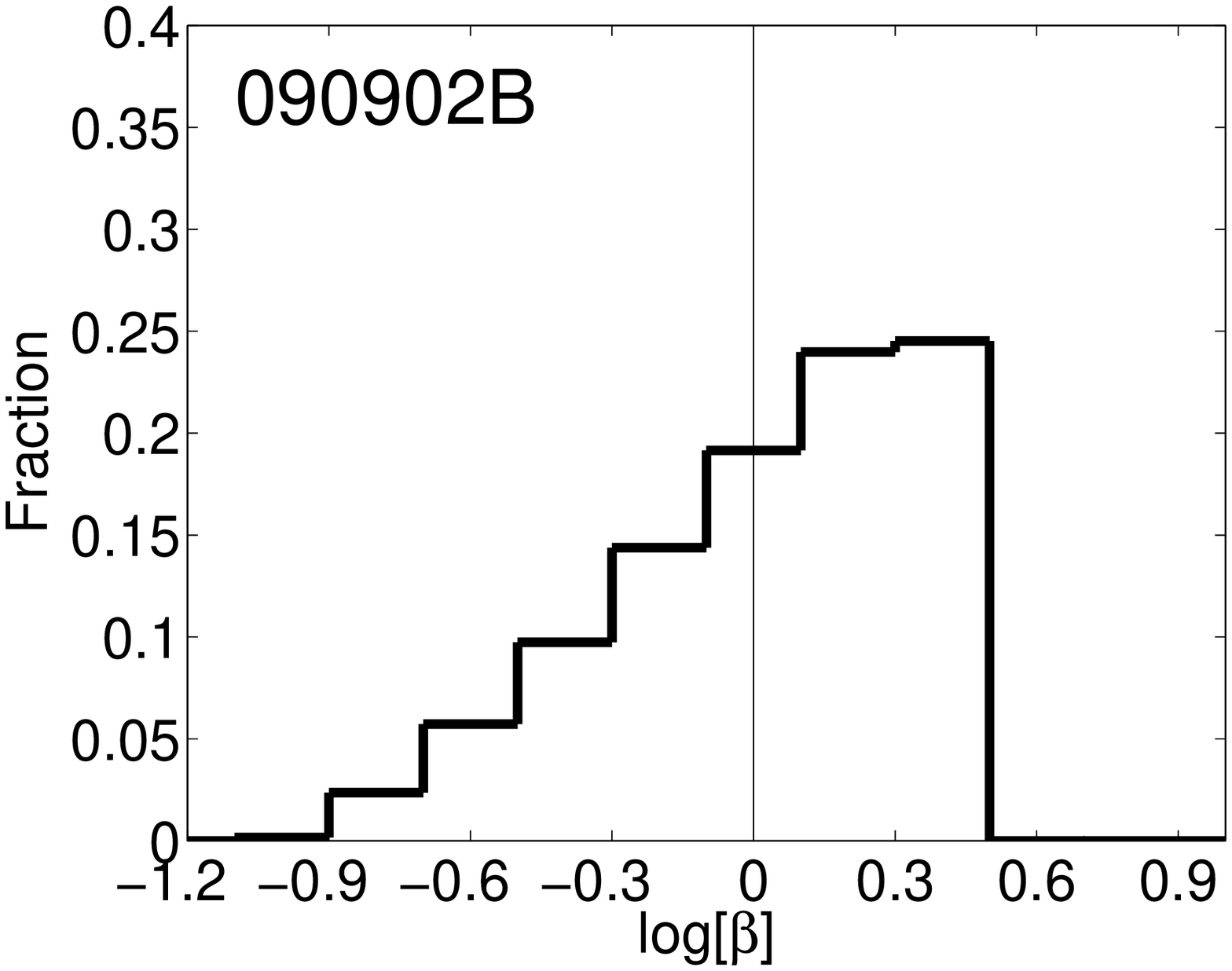} 
\caption{Normalized histograms of inferred expansion velocity,
$\beta\equiv v/c$, at the time of our observations.  A value of
$\lesssim 1$ (vertical lines) is required for self-consistency, and
this is indeed the case for the bulk of the acceptable solutions.
Note that the scales for the three groups are different.}
\label{fig:beta} 
\end{figure}

The resulting solutions for each burst can be cast in terms of a
two-dimensional parameter space in $n_e$ versus $E_{\rm ST}$.  Thus,
there is a degeneracy between the two parameters, in the sense that
larger densities lead to lower energies.  Clearly, the bursts in Group
A, for which the peak of the synchrotron spectrum is well-determined,
lead to the best constraints in this two-dimensional phase space.
Indeed, as shown for GRB\,980703 (Figure~\ref{fig:examples}), the
allowed range of energies for all solutions that reproduce the
observed flux density is about $10^{48}-10^{53}$ erg, while the
solutions that also satisfy the requirements that $(E_B + E_e)\lesssim
E_{\rm ST}/2$ span a much narrower range of about $10^{51}-10^{52}$
erg, with a roughly log-normal distribution centered on ${\rm
log}(E_{\rm ST})\approx 51.6$.

The bursts with only single-frequency observations (Group B) cover a
much larger area in the $E_{\rm ST}-n_e$ phase-space since only an
upper bound can be placed on $\nu_m$ and $\nu_a$.  To further
constrain the energy we place an additional conservative constraint on
the density of $n_e<100$ cm$^{-3}$, motivated by the results of
detailed broad-band modeling that show $n_e\sim 0.1-10$ cm$^{-3}$
(e.g., \citealt{pk01,pk02,yhs+03}).  Given the anti-correlation
between density and energy, our conservative limits lead to a wider
range of allowed energies than if we chose a limit of $n_e<10$
cm$^{-3}$.  An example of this additional constraint for a Group B
burst (GRB\,010921) is shown in Figure~\ref{fig:examples}.  We do not
place a lower bound on the density, since for the phase-space of
allowed solutions this would not lead to a significant change in the
energy distribution.  We stress that beyond placing an upper bound on
$n_e$ no constraints have been placed on the distributions of either
$n_e$ or $\beta$ since both are inferred, and not input, parameters in
our model.

As noted above, we do not consider the energies for the bursts in
Group C since the upper limits generally allow a wide range of
solutions that are not consistent with the Sedov-Taylor formulation.
This indicates that the limits are generally not deep enough to
provide a meaningful constraint on the energy.  Future deep radio
observations may provide much better constraints (see below).

The resulting energy probability distributions for the bursts in Group
A and Group B are shown in Figure~\ref{fig:hist}.  The median energy
and $90\%$ confidence range (i.e., $5-95\%$ of the distribution) for
each burst are listed in Table~\ref{tab:energy}.  We include in these
ranges the small subset of solutions that lead to $\beta$ values in
slight excess of 1 since these are at most mildly relativistic and
furthermore do not significantly change the distributions
(Figure~\ref{fig:hist}).  We find that varying the electron power law
index over the range $p=2.1-2.5$ (e.g., \citealt{cep+10}) leads to a
change in the median energy of only $0.1-0.2$ dex (compared to our
fiducial value of $p=2.2$), with larger values of $p$ leading to lower
median energies.  Similarly, varying the magnetic energy fraction away
from equipartition to $\epsilon_B=0.1$ and $0.01$, leads to an increae
in the median energy of about 0.25 and 0.5 dex, respectively.  Both of
these effects are much smaller than the overall spread in energy for
each burst, but they do produce minor systematic trends.

The combined distribution for the subset of 11 bursts whose solutions
are generally self-consistent is shown in Figure~\ref{fig:eall}.  The
median and $90\%$ confidence ranges are $7\times 10^{51}$ erg and
$1.1\times 10^{50}-3.3\times 10^{53}$ erg.

\section{Discussion and Conclusions}
\label{sec:comp}

The key results of our analysis are that the median energy for the 11
bursts with self-consistent solutions is $E_K\approx 7\times 10^{51}$
erg, while the $90\%$ confidence range is $1.1\times 10^{50}-3.3\times
10^{53}$ erg.  The median value is about a factor of 3 times higher
than previous calorimetric measurements for GRBs 970508, 980703, and
030329, for which energies of $3\times 10^{51}$, $3\times 10^{51}$,
and $10^{51}$ erg, respectively, were determined \citep{bkf04,fsk+05}.

\begin{figure}
\includegraphics[width=1.55in]{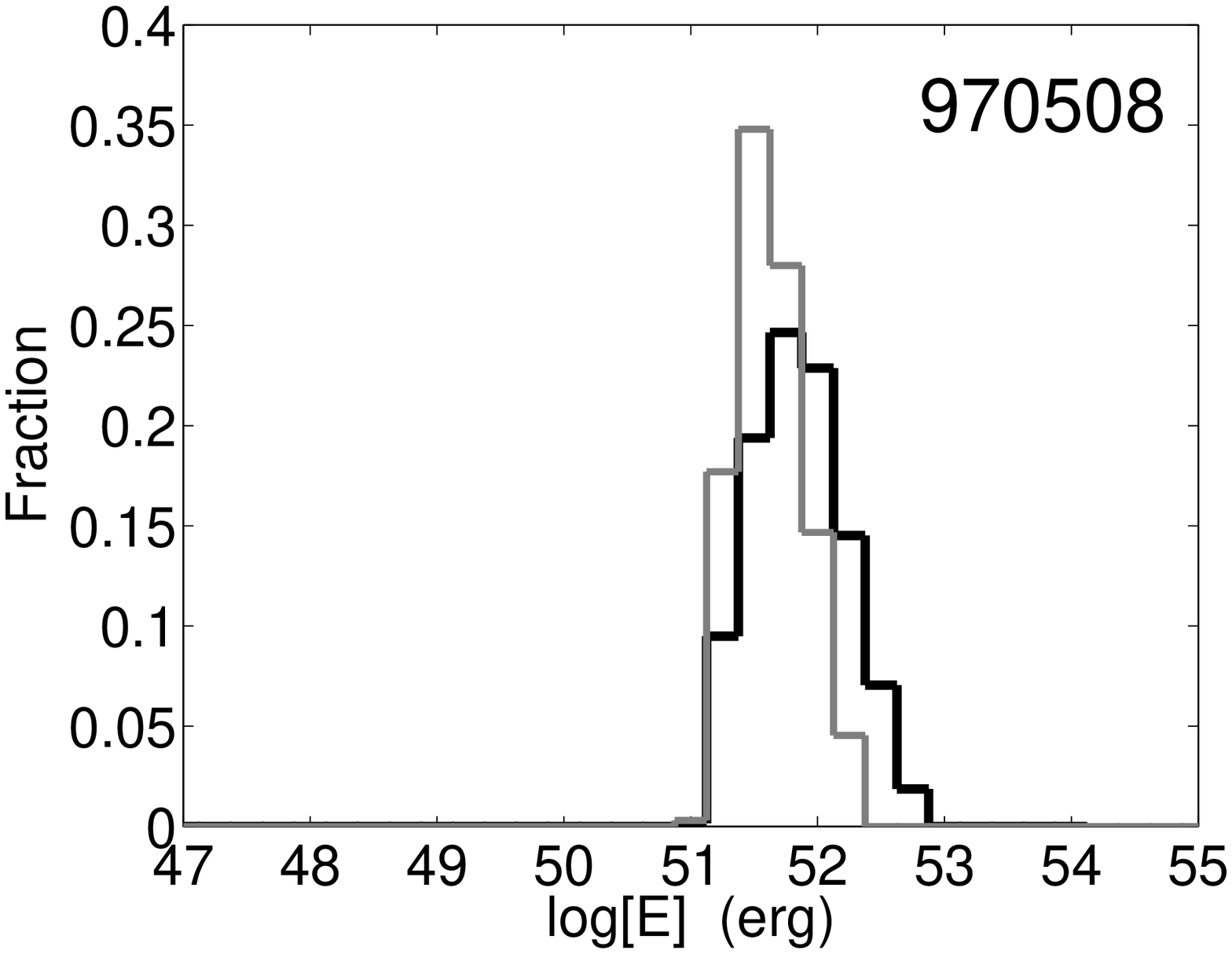} 
\includegraphics[width=1.55in]{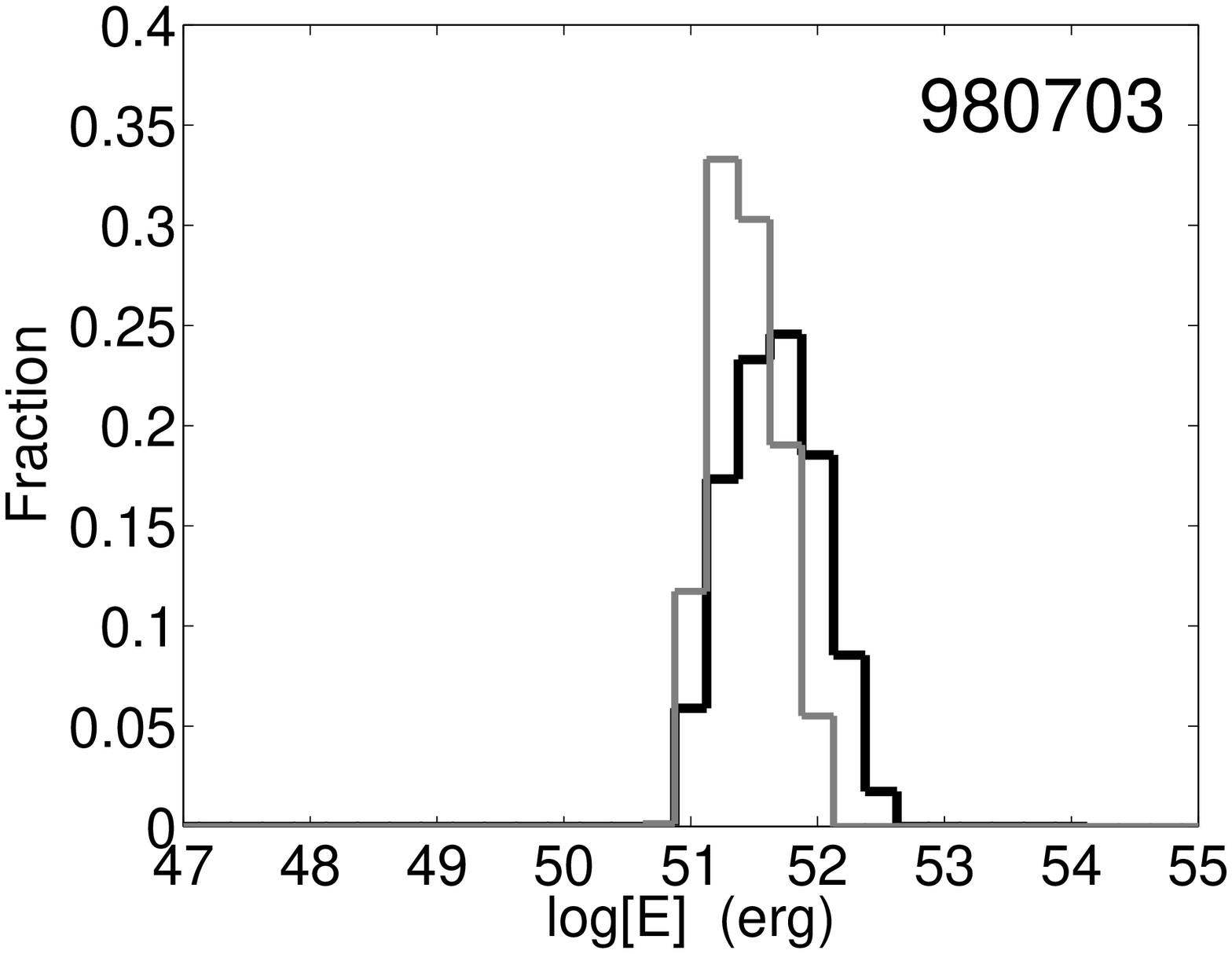} 
\includegraphics[width=1.55in]{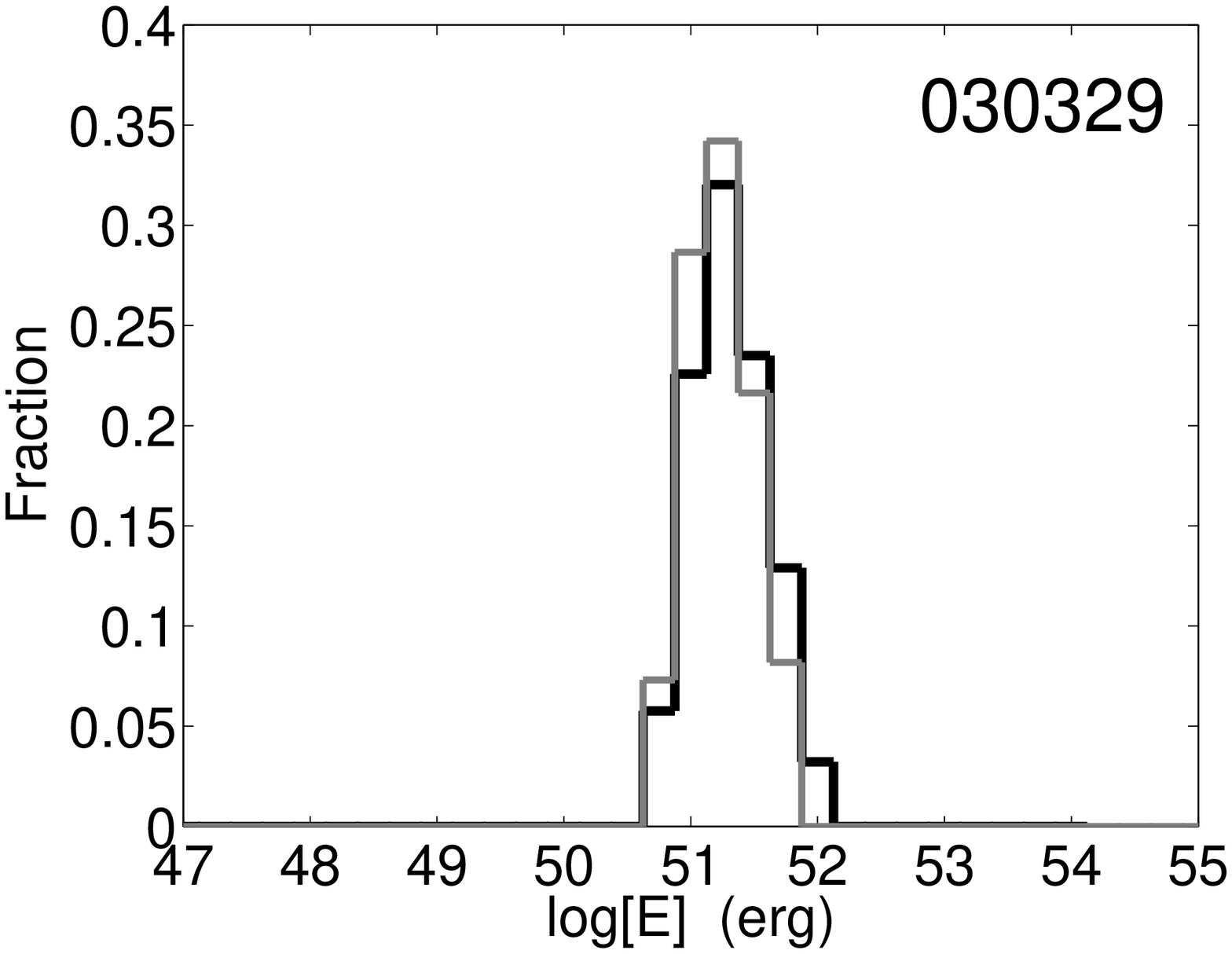} 
\includegraphics[width=1.55in]{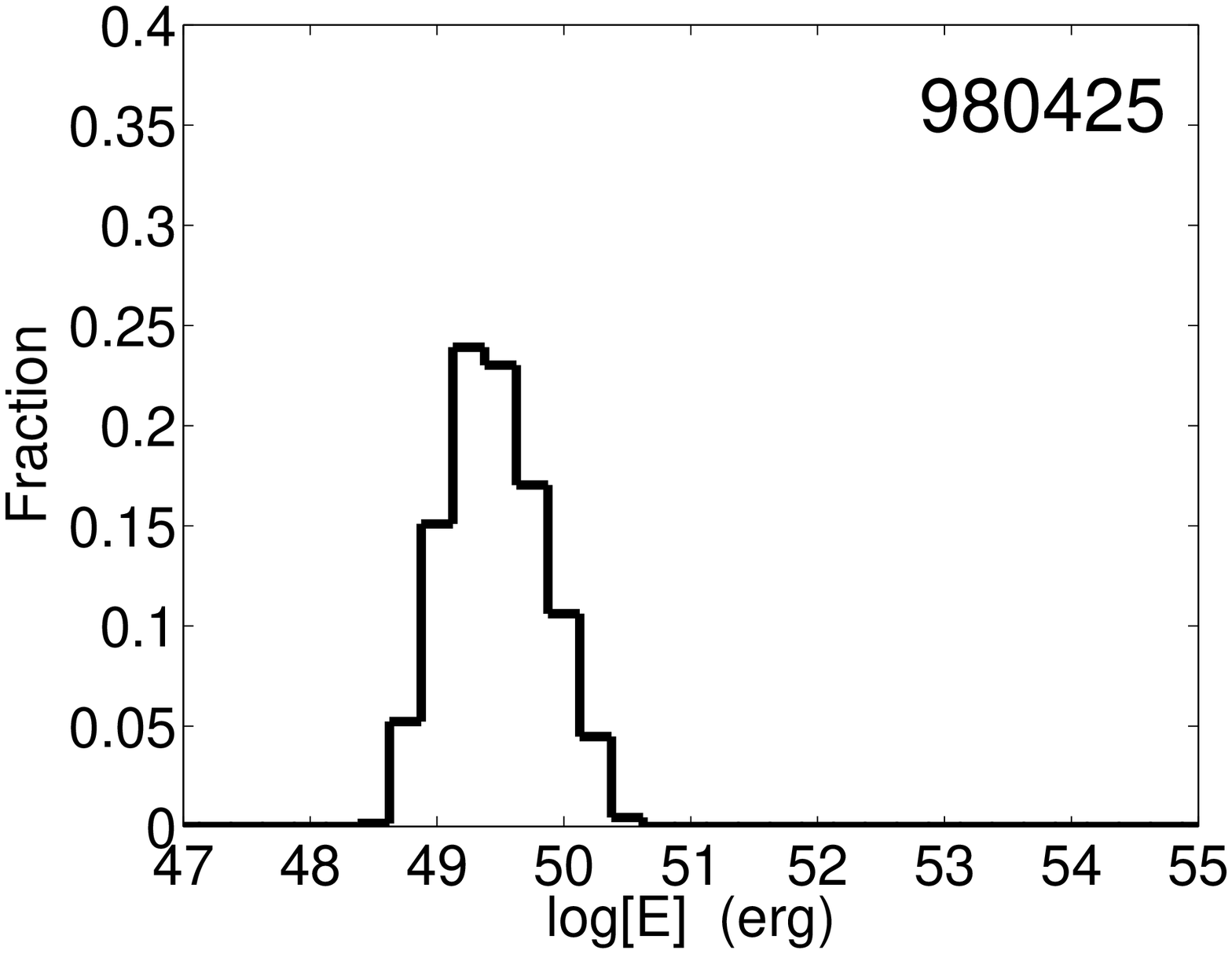} \\
\includegraphics[width=1.55in]{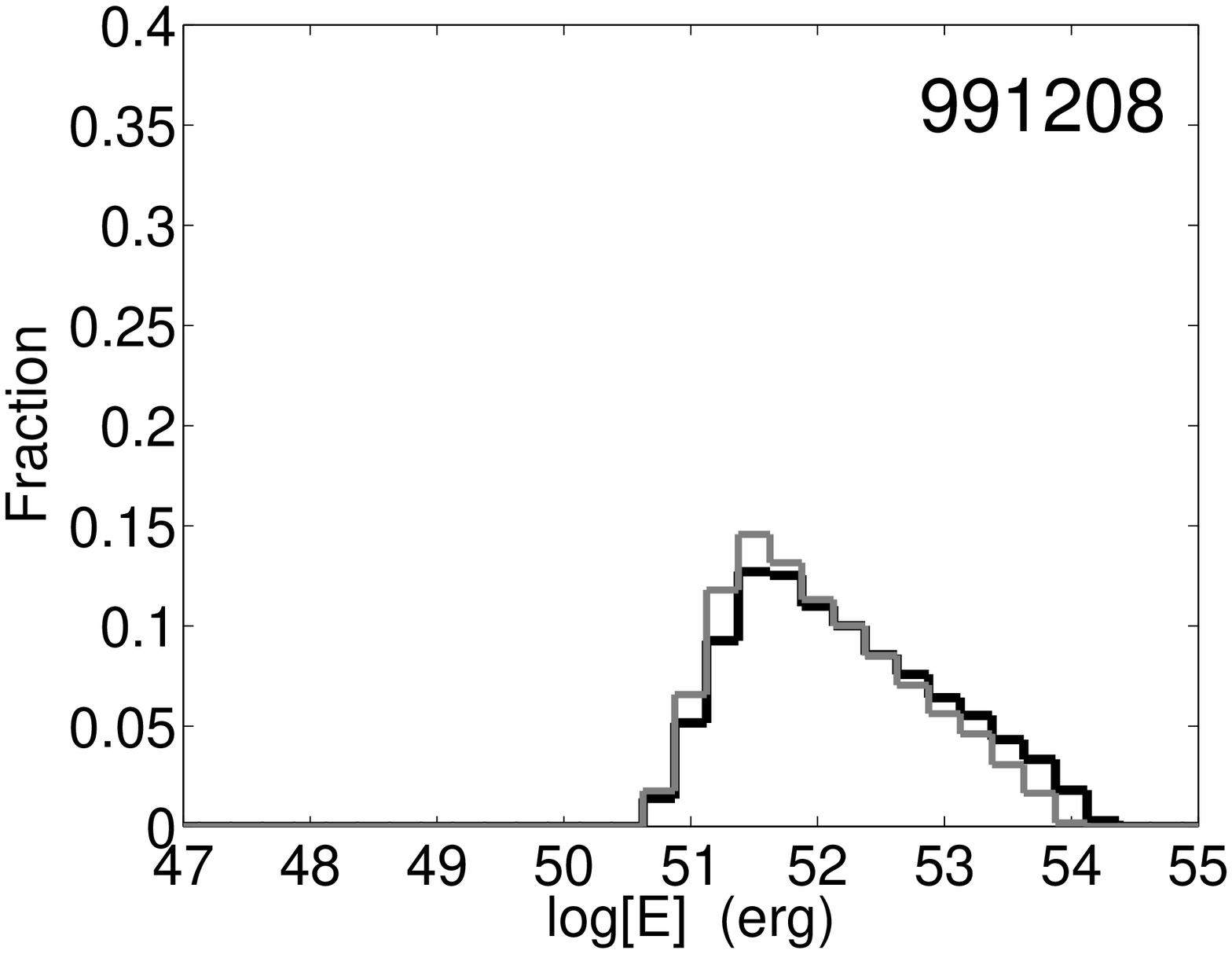} 
\includegraphics[width=1.55in]{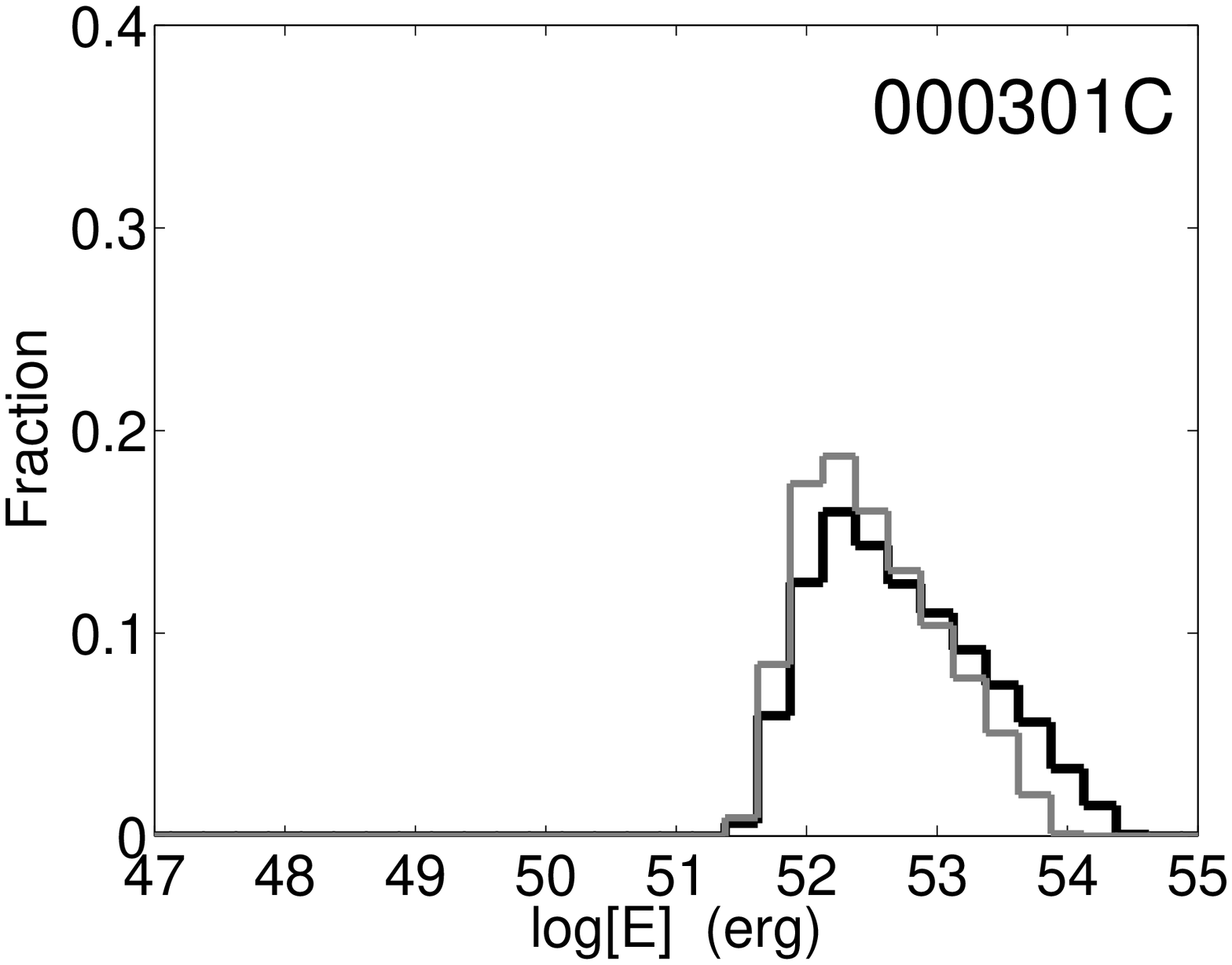} 
\includegraphics[width=1.55in]{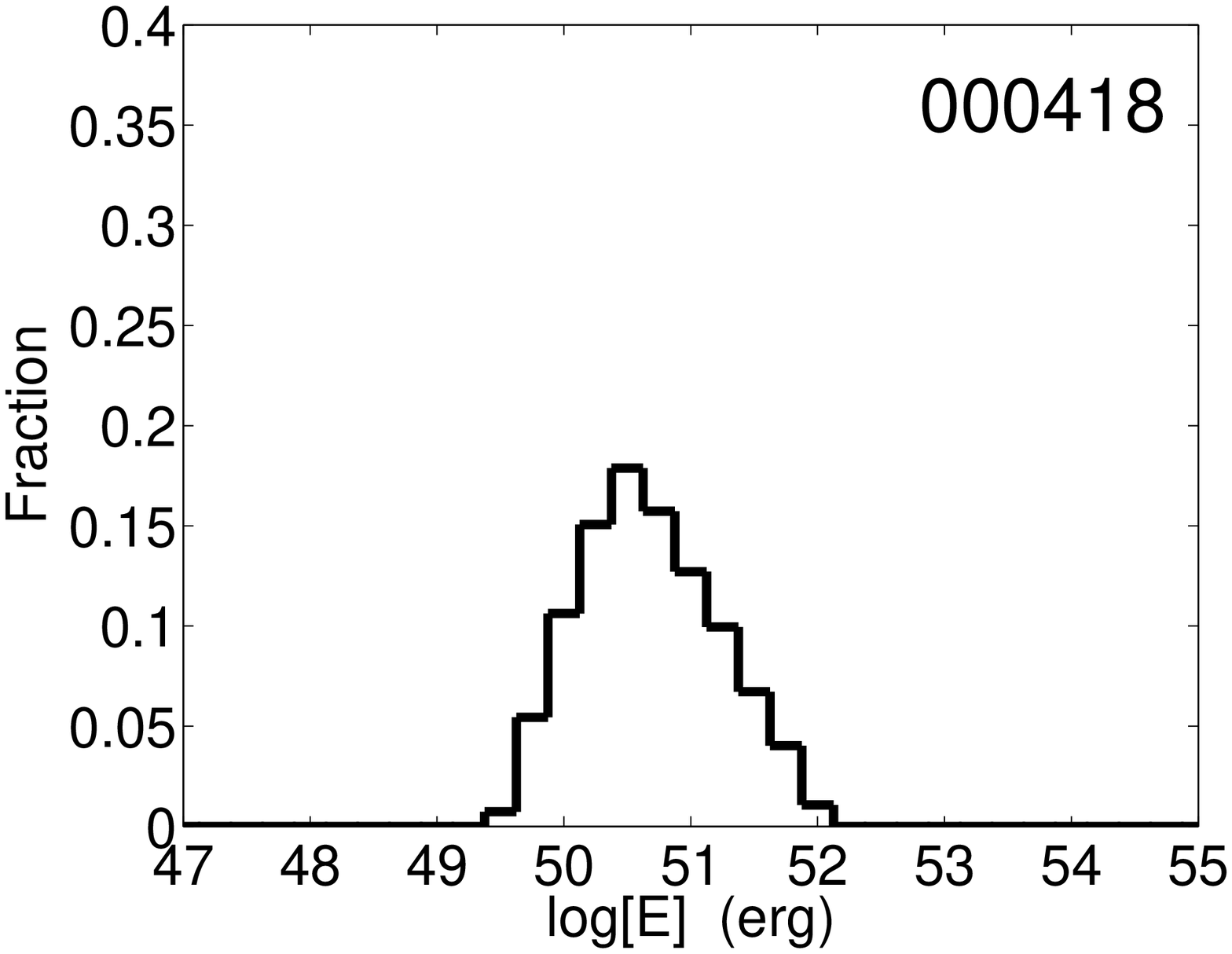} 
\includegraphics[width=1.55in]{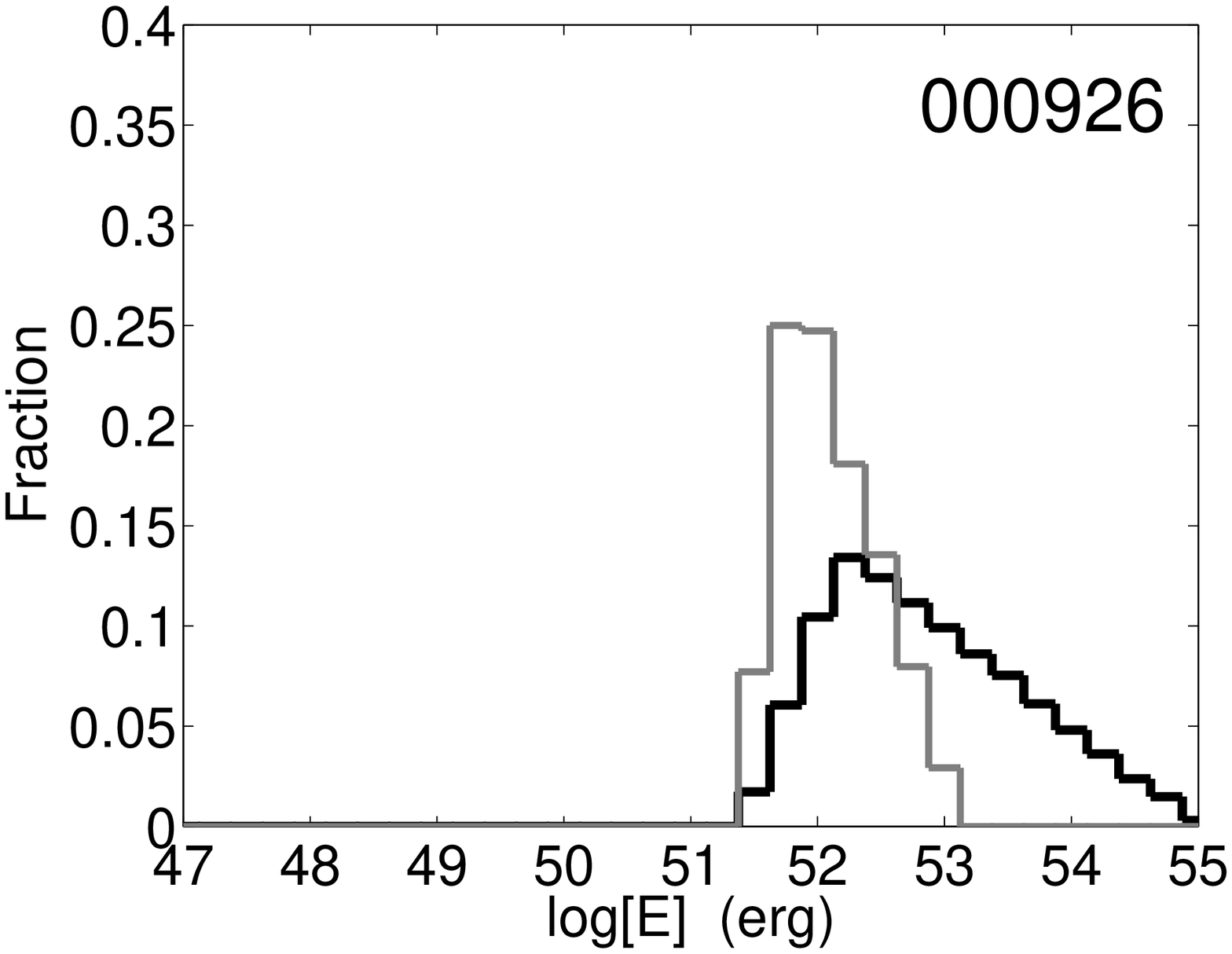} \\
\includegraphics[width=1.55in]{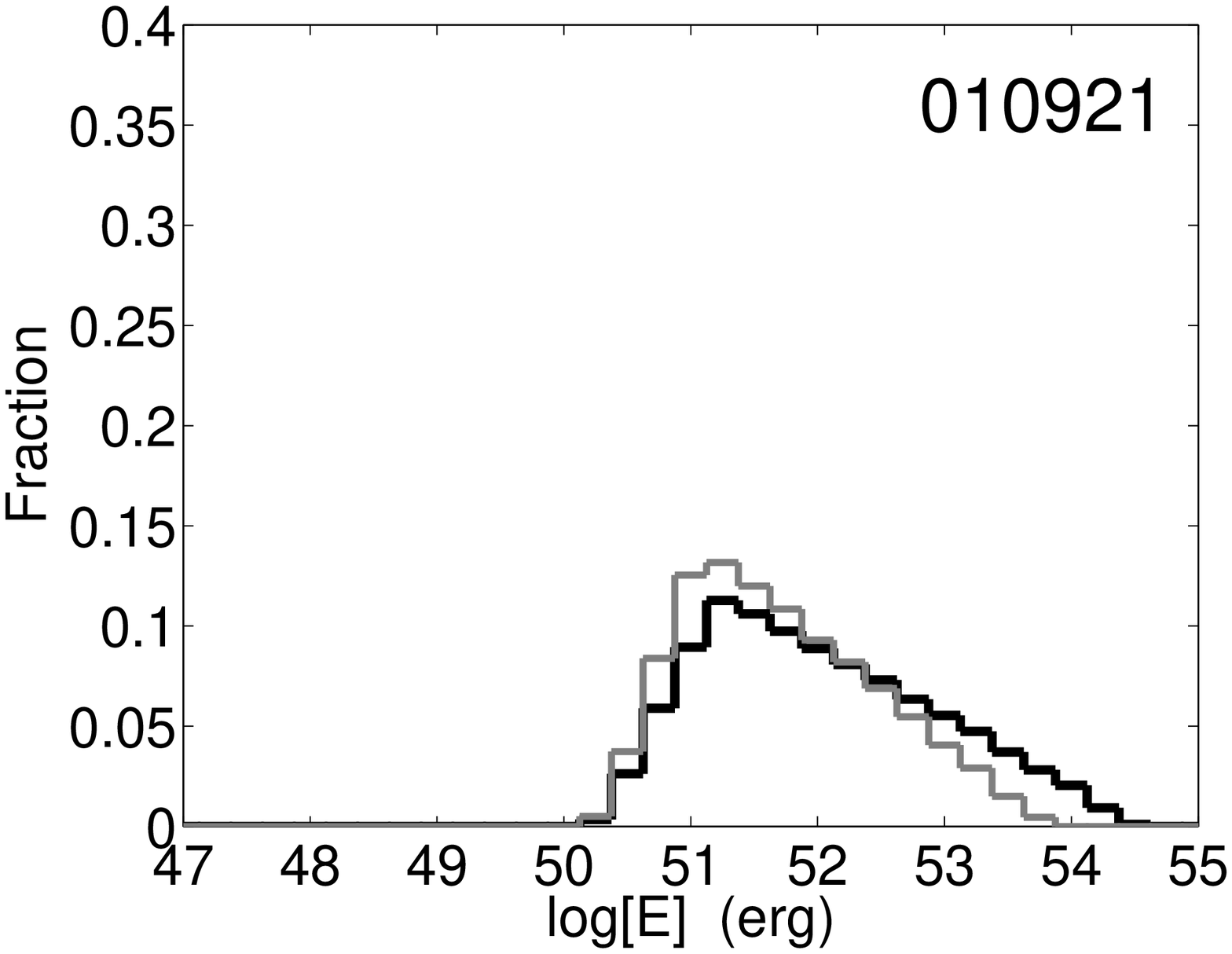} 
\includegraphics[width=1.55in]{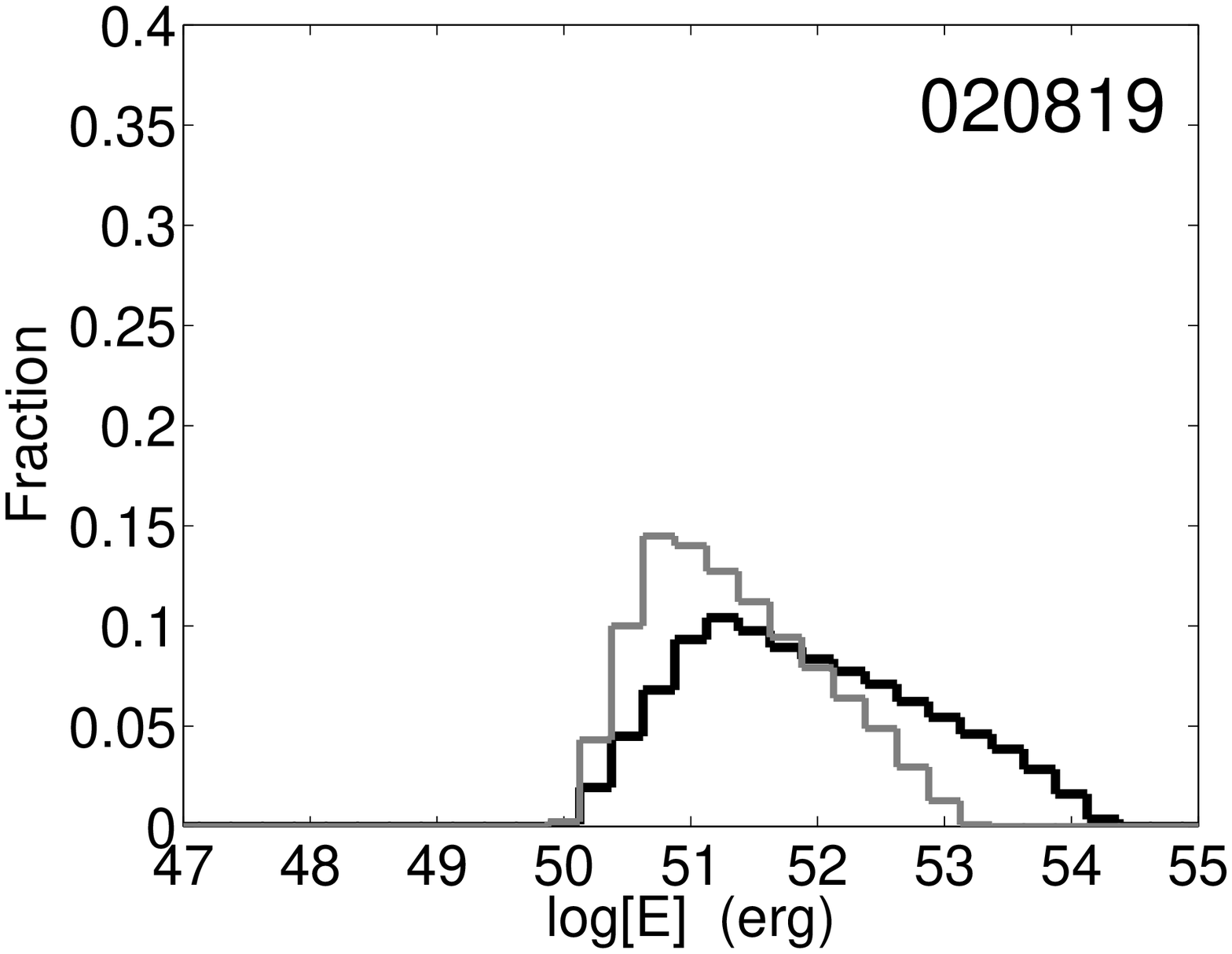} 
\includegraphics[width=1.55in]{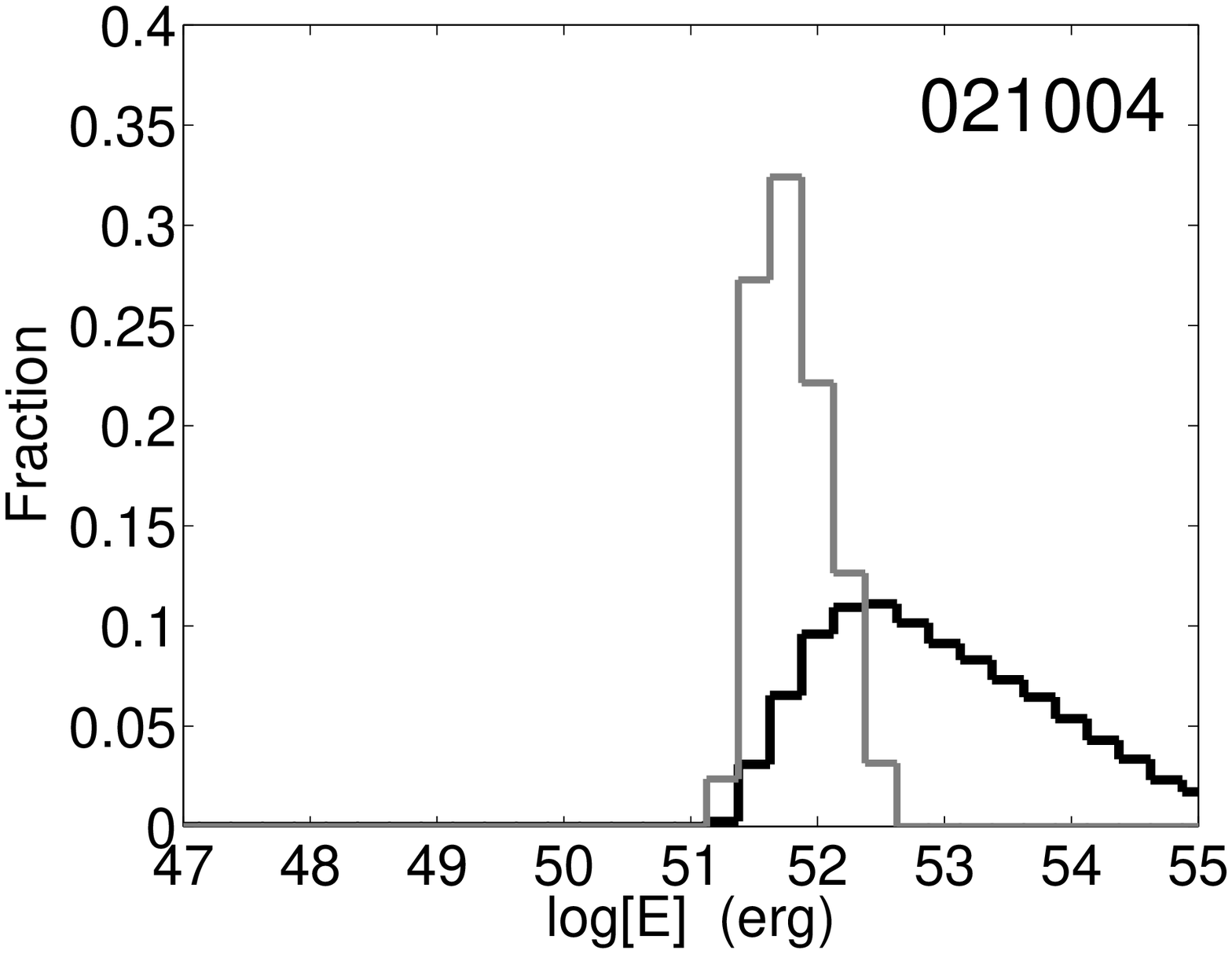} 
\includegraphics[width=1.55in]{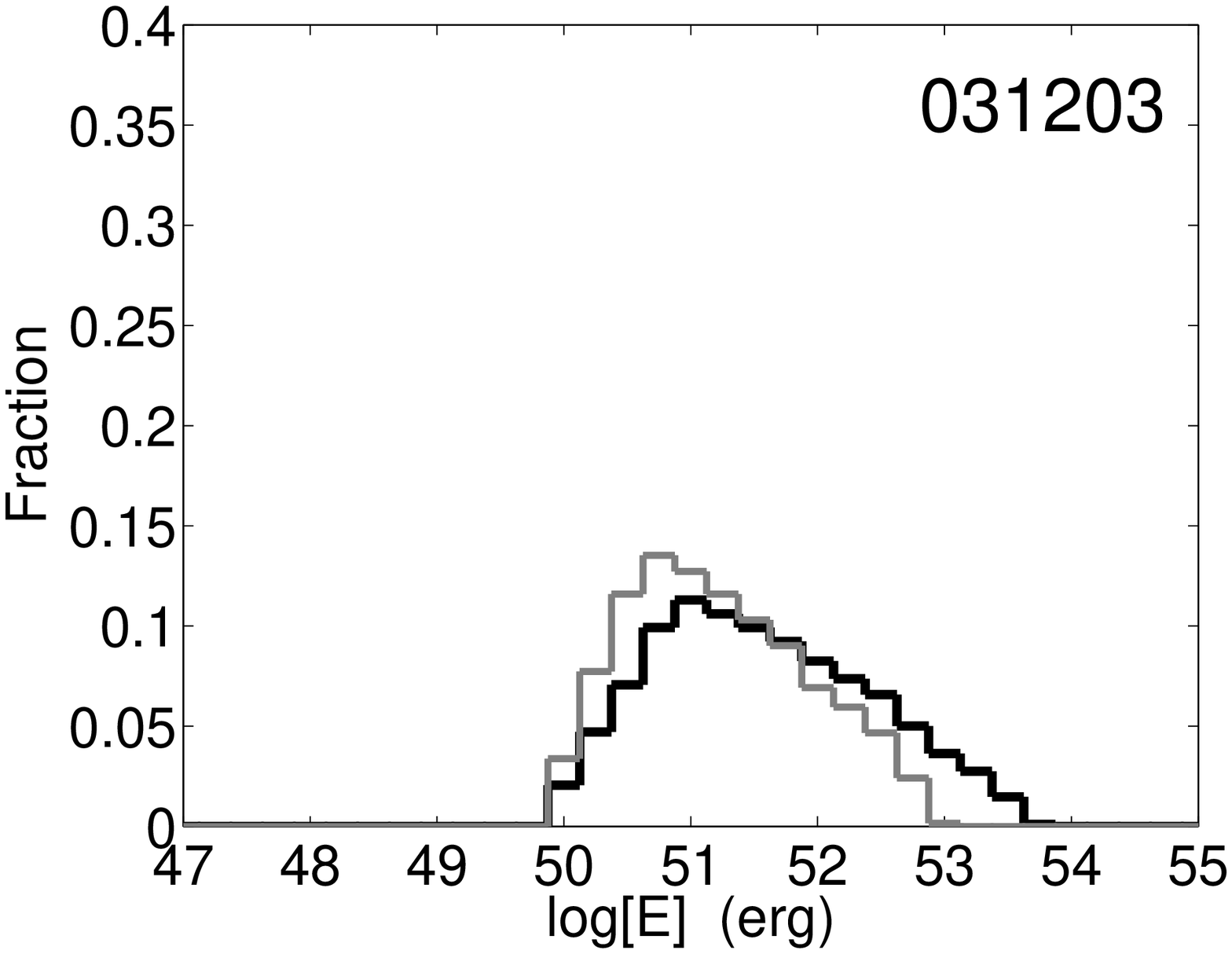} \\
\includegraphics[width=1.55in]{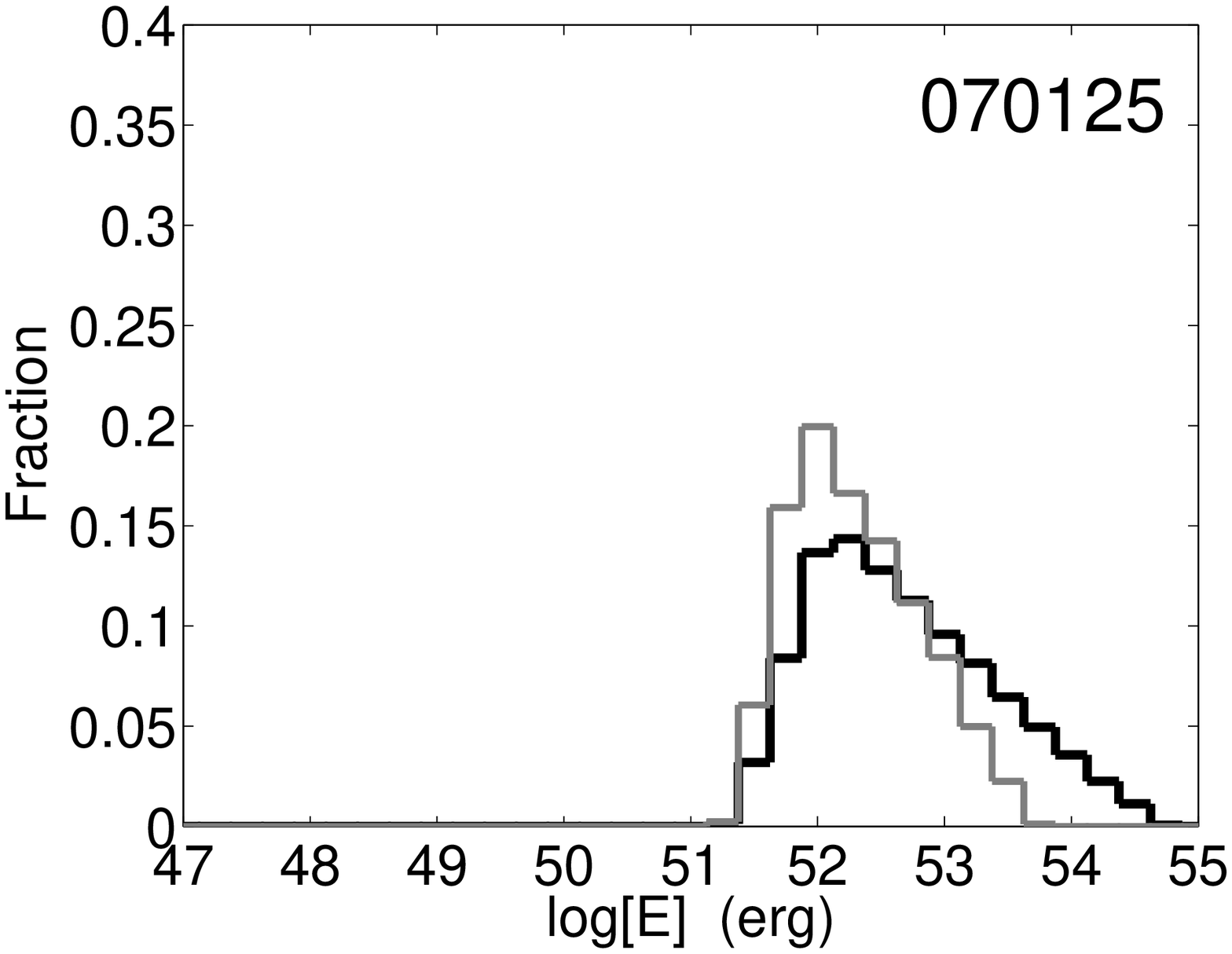} 
\includegraphics[width=1.55in]{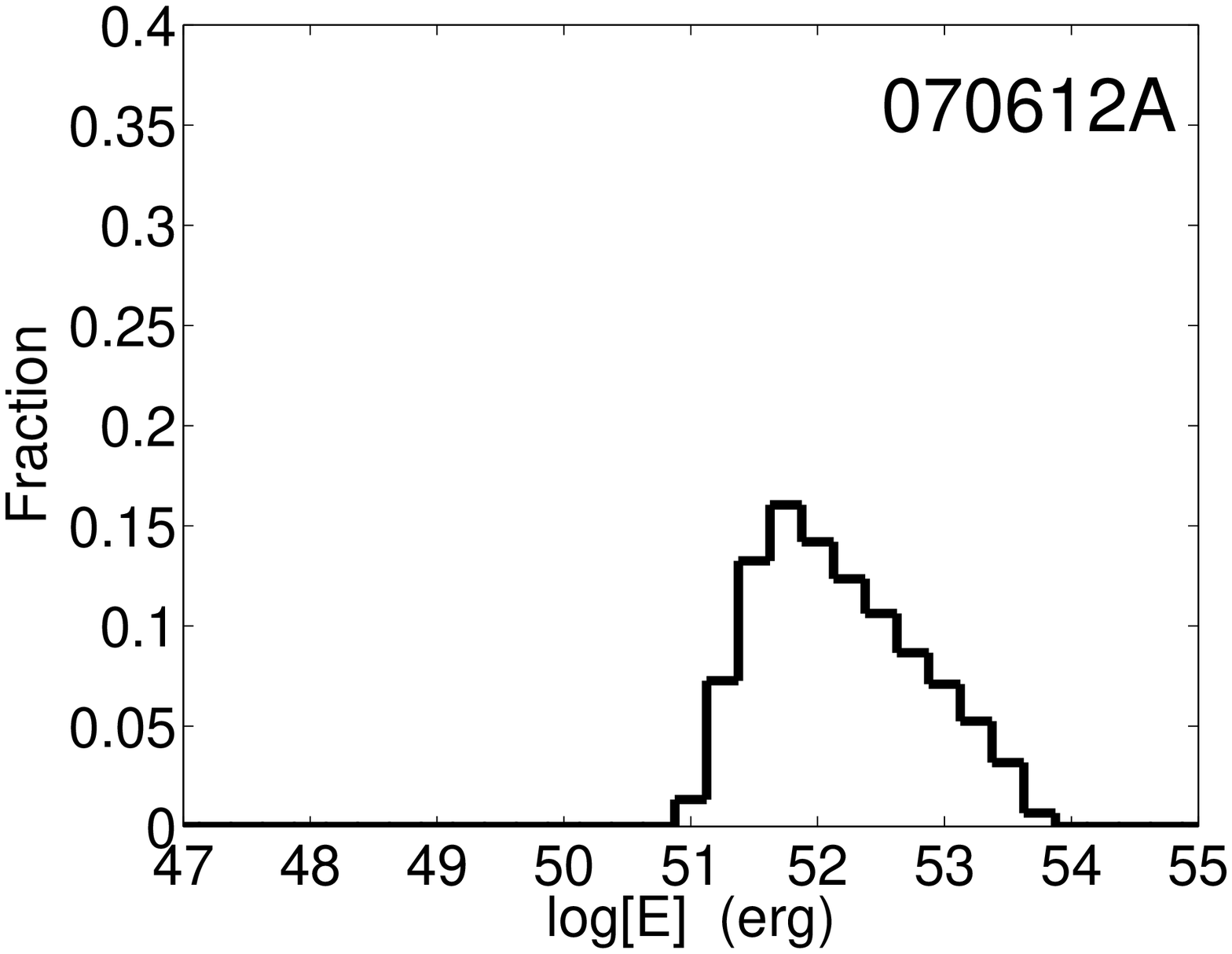} 
\caption{Normalized histograms of GRB energies calculated using the
Sedov-Taylor solution.  The light gray histograms for some Group B
bursts indicate the subset of solutions with a strict cut-off of
$\beta<1$.  GRBs 000926, 020819, and 021004 are rejected from our
sample since the bulk of their solutions still lead to relativistic
expansion.}
\label{fig:hist}
\end{figure}

\begin{figure}
\epsscale{1}
\plotone{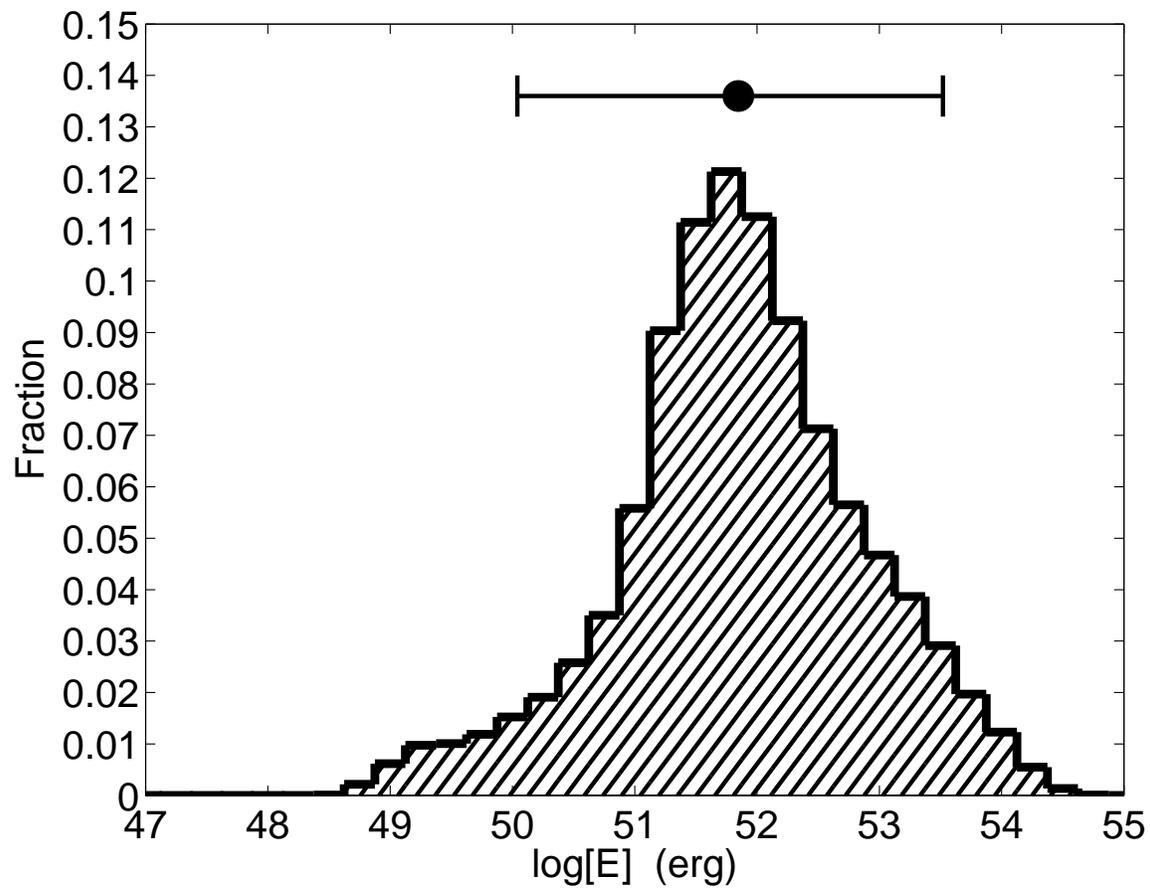}
\caption{Normalized distribution of GRB kinetic energies calculated
using the Sedov-Taylor solution for the sample of 11 bursts with
self-consistent solutions (Figure~\ref{fig:hist}).  The median and
90\% confidence range are marked by a horizontal bar.}
\label{fig:eall} 
\end{figure}

\begin{figure}
\epsscale{1}
\plotone{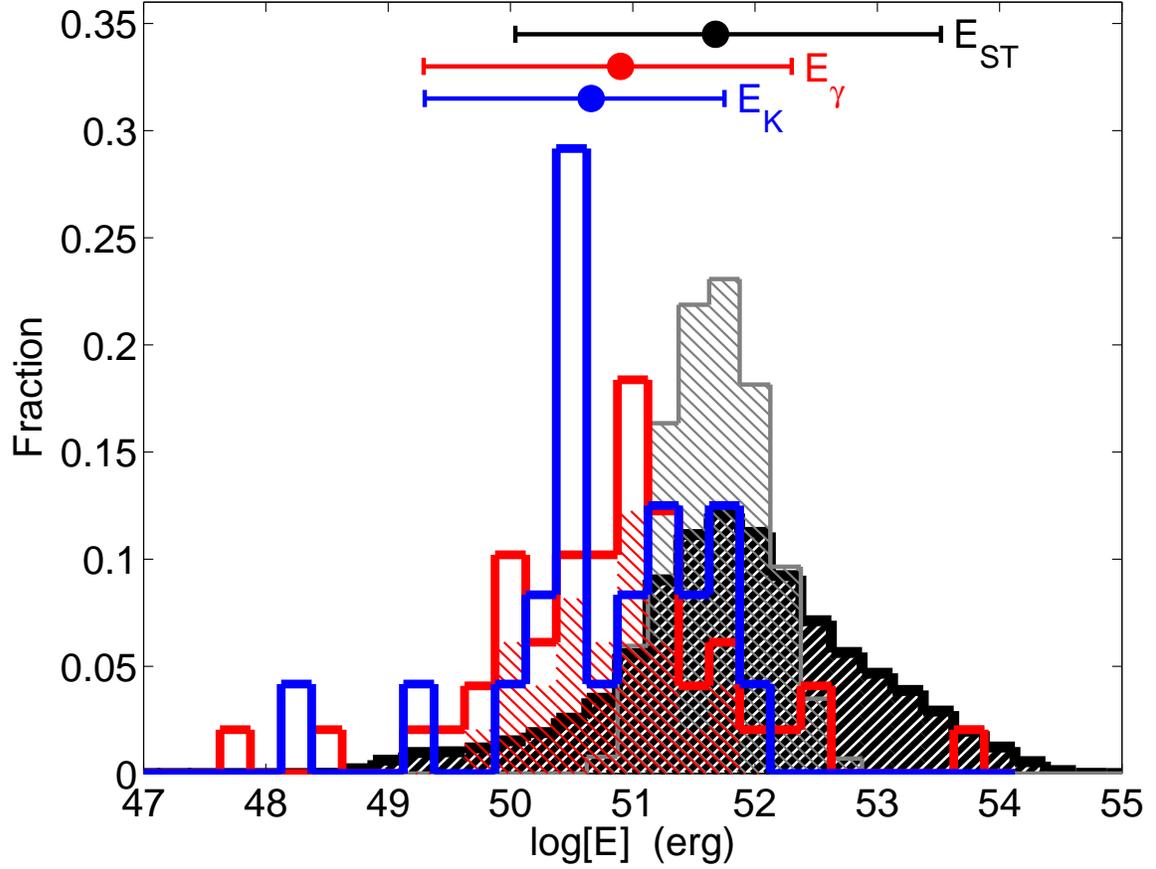}
\caption{Normalized distributions of GRB kinetic energies calculated
using the Sedov-Taylor solution (black), and for the subset of 3
bursts in Group A (gray).  Also shown for comparison are the
distributions of beaming-corrected $\gamma$-ray energies (red: hatch =
known $\theta_j$ values; open = $\theta_j$ lower or upper limits;
\citealt{fb05}) and beaming-corrected kinetic energies from broad-band
early afterglow modeling (blue;
\citealt{pk02,bkp+03,yhs+03,skb+04,skb+04b,skn+06,cfh+10,cfh+10b}).
The median and $90\%$ confidence range for each energy component are
marked by a horizontal bar.  Our inferred median energy and $90\%$
confidence range are larger than the median of both $E_\gamma$ and
$E_K$, but this is mainly due to the bursts in Group B for which the
spectral peak is not measured.  Future observations with the EVLA will
lead to much tighter constraints (see gray histogram) for a larger
sample.}
\label{fig:ecomp}
\end{figure}

Similarly, the inferred energies are somewhat larger than the
distributions of beaming-corrected $\gamma$-ray and kinetic energies
inferred from broad-band multi-epoch studies (Figure~\ref{fig:ecomp}).
From various such analyses, the median $\gamma$-ray energy is $\langle
E_\gamma\rangle \approx 8\times 10^{50}$ erg
\citep{fks+01,bfk03,fb05}, while the median kinetic energy is $\langle
E_K\rangle\approx 5\times 10^{50}$ erg (e.g.,
\citealt{pk01,pk02,yhs+03}); see Figure~\ref{fig:ecomp}.  In both
cases the $90\%$ range spans about 2.5 orders of magnitude, somewhat
narrower than our inferred $90\%$ confidence range for $E_{\rm ST}$.
The extension to larger energies found in our analysis mainly reflects
the lack of spectral peak determinations for the bursts with
single-frequency observations (see Figure~\ref{fig:hist}).  These
large energies can be generally eliminated with a measurement of the
synchrotron peak in the GHz frequency range (e.g., Group A bursts;
Figure~\ref{fig:ecomp}).

In the context of our results we note that recent numerical work by
\citet{zm09} led these authors to conclude that the timescale to reach
isotropy is $\sim 10^2$ yr rather than $\sim 1$ yr as indicated by the
analytic formulation of \citet{lw00} which we follow here.  As a
result, they note that using the Sedov-Taylor formulation may lead to
an erroneous estimate of the kinetic energy.  However, inspection of
the resulting potential disrepancies reveals that this effect is at
most a factor of 2 {\it as long as self-consistency between the
inferred energy and density and the transition to the Sedov-Taylor
phase is ensured} (see their Figure 10).  The discrepancies become
larger if the wrong timescale is assumed for the transition to
non-relativistic expansion, but this quantity is not a free parameter.
Indeed, our distributions of $\beta$ values point to self-consistency
for most bursts, and allow us to reject objects that are potentially
still relativistic.  Since the potential systematic uncertainty of
about a factor of 2 is significantly smaller than the overall spread
in allowed energy for each burst, we do not consider this to be an
obstacle to our analysis, or to future work on the energy scale using
late-time radio measurements.

As clearly demonstrated in Figure~\ref{fig:hist}, the most constrained
energy determinations require a measurement of the synchrotron
spectral peak (Group A); the absence of such a constraint requires
additional assumptions about the circumburst density and results in a
much wider energy range.  Indeed, this is the key reason for the wider
range of allowed high energy solutions ($\gtrsim 10^{52}$ erg)
compared to the results for $E_\gamma$ and $E_K$
(Figure~\ref{fig:ecomp}).  Observations of GRBs 970508, 980703, and
030329 demonstrate that the spectral peak is typically located at
$\sim {\rm few}$ GHz on a timescale of $\sim 150$ d.  Thus,
observations in the $1-10$ GHz range on a timescale of $\sim {\rm
few}$ hundred days should allow us to determine the peak flux and
frequency.  This will in turn provide an energy estimate with a
similar level of precision to the results of early-time broad-band
modeling.

This is a fortuitous conclusion since with the full frequency coverage
of the Expanded VLA (EVLA) it will soon be possible to cover this
entire range in a few hours of observations to a sensitivity that is
about an order of magnitude better than the VLA.  As we demonstrated
here, such a modest investment of observing time ($2-3$ hours per
burst) can yield a robust estimate of the GRB energy distribution,
{\it regardless} of the ability to measure jet opening angles.
Pursuing these observations for all bursts with a measured redshift
will require only $\sim 50-100$ hr of EVLA time per year.  Indeed,
with such observations we should be able to constrain the energy
distribution to a comparable level as existing studies within a single
year given that about 30 GRBs with known redshifts just from 2009 are
now available for EVLA observations (a similar sample is available
from 2008 bursts).  In the longer term, the large number of objects
will allow us to test the energy distribution as a function of
redshift, at least over the range $z\sim 1-3$ where the bulk of the
detected bursts occur \citep{bkf+05,jlf+06}.  Similarly, this approach
will be particularly useful for bursts that lack detailed optical or
X-ray light curves due to observational constraints or dust
extinction, and for bursts with controversial estimates of the jet
opening angles.

\acknowledgements We thank Dale Frail and Eli Waxman for helpful
discussions and comments on the manuscript.  


\clearpage
\begin{deluxetable}{lcccll}
\tabletypesize{\scriptsize}
\tablecolumns{6}
\tabcolsep0.08in\footnotesize
\tablewidth{0pc}
\tablecaption{Late-time Radio Afterglow Measurements
\label{tab:data}}
\tablehead{
\colhead{GRB}        &
\colhead{$z$}        &
\colhead{$\delta t$} &
\colhead{$\nu$}      &
\colhead{$F_\nu\,^a$}&
\colhead{Ref.}       \\
\colhead{}           &
\colhead{}           &
\colhead{(d)}        &
\colhead{(GHz)}      &
\colhead{($\mu$Jy)}  &
\colhead{}           
}
\startdata
970508  & 0.835	 & 117.55 & 8.46 & $355\pm 47$	  & \citet{fwk00} \\
    	&        & 117.55 & 4.86 & $425\pm 57$	  & \\
    	&        & 117.55 & 1.43 & $206\pm 63$	  & \\
970828  & 0.958  & 157.99 & 8.46 & $<51$          & \citet{dfk+01} \\
980425  & 0.0085 & 248.20 & 8.70 & $700\pm 200$   & \citet{kfw+98} \\
980703  & 0.966	 & 143.79 & 8.46 & $110\pm 20$	  & \citet{fyb+03} \\
        &	 & 143.79 & 4.86 & $146\pm 24$	  & \\
        &	 & 134.85 & 1.43 & $99\pm 25$	  & \\
990506  & 1.307  & 141.23 & 8.46 & $<75$          & \citet{tbf+00} \\ 
991208  & 0.706	 & 291.58 & 8.46 & $51\pm 15$	  & \citet{gfs+03} \\
000210  & 0.846  & 108.37 & 8.46 & $<78$          & \citet{fkb+03} \\
000301C & 2.030	 & 506.10 & 8.46 & $39\pm 11$	  & \citet{bsf+00} \\
000418  & 1.118	 & 405.76 & 8.46 & $38\pm 11$	  & \citet{bdf+01} \\
000911  & 1.058  & 125.78 & 8.46 & $<54$          & \citet{pbk+02} \\
000926  & 2.066	 & 257.43 & 8.46 & $75\pm 21$	  & \citet{hys+01} \\
010222  & 1.477  & 206.63 & 8.46 & $<42$          & \citet{fkb+03} \\
010921  & 0.451  & 225.42 & 8.46 & $52\pm 15$     & \citet{fkb+03} \\
011121  & 0.362  & 132.07 & 8.70 & $<141$         & \citet{fkb+03} \\
020819  & 0.411	 & 126.45 & 8.46 & $79\pm 25$	  & \citet{jff+05} \\
021004  & 2.329	 & 140.24 & 8.46 & $94\pm 16$	  & \citet{fkb+03} \\
030226  & 1.986	 & 113.85 & 1.43 & $<117$         & \citet{fkb+03} \\
030329  & 0.168  & 135.48 & 8.46 & $1525\pm 56$	  & \citet{fsk+05} \\
    	&        & 129.57 & 4.86 & $1955\pm 62$	  & \\
    	&        & 129.58 & 1.43 & $1276\pm 56$	  & \\
031203  & 0.105	 & 137.15 & 8.46 & $426\pm 37$ 	  & \citet{skb+04} \\
050416A & 0.654  & 182.28 & 8.46 & $<114$         & \citet{snc+07} \\
070125  & 1.547	 & 341.96 & 8.46 & $64\pm 18$	  & \citet{ccf+08} \\
070612A & 0.617	 & 488.54 & 8.46 & $101\pm 39$	  & \citet{fkb+03} \\
090323  & 3.570	 & 131.18 & 8.46 & $<81$          & \citet{cfh+10b}\\
090902B & 1.822	 & 199.16 & 8.46 & $<48$          & \citet{cfh+10b}
\enddata
\tablecomments{$^a$ Limits are $3\sigma$.}
\end{deluxetable}

\clearpage
\begin{deluxetable}{lcccc}
\tabletypesize{\scriptsize}
\tablecolumns{5}
\tabcolsep0.08in\footnotesize
\tablewidth{0pc}
\tablecaption{GRB Energies Inferred from Calorimetry
\label{tab:energy}}
\tablehead{
\colhead{GRB}                                   & 
\colhead{$\langle{\rm log}(E_{\rm ST})\rangle$} &
\colhead{${\rm log}(E_{\rm ST})_{,90}\,^a$}     &
\colhead{${\rm log}(E_\gamma)\,^b$}             &
\colhead{${\rm log}(E_K)\,^c$}                  \\
\colhead{}                                      &
\colhead{(erg)}                                 &
\colhead{(erg)}                                 &
\colhead{(erg)}                                 &
\colhead{(erg)}                                  
}
\startdata
\multicolumn{5}{c}{\bf{Group A}}  \\
\hline 
970508  &  51.8  & $51.3 - 52.5$ & 50.6 & 51.3 \\
980703  &  51.6  & $51.1 - 52.2$ & 51.0 & 51.5 \\
030329  &  51.3  & $50.9 - 51.8$ & 49.9 & 50.4 \\\hline
\multicolumn{5}{c}{\bf{Group B}}  \\
\hline
980425  &  49.4  & $48.9 - 50.1$ & 47.8    & $\sim 50$ \\
991208$\,^d$ &  51.9  & $51.0 - 53.4$ & 51.2    & 50.4 \\
        &  52.1  & $51.1 - 53.6$ &         &      \\
000301C &  52.4  & $51.8 - 53.5$ & 50.9    & 50.5 \\
        &  52.6  & $51.8 - 53.9$ &         &      \\
000418  &  50.6  & $49.8 - 51.6$ & 51.7    & 51.5 \\
000926  &  52.0  & $51.6 - 52.8$ & 51.2    & 51.2 \\
        &  52.7  & $51.8 - 54.3$ &         &      \\
010921  &  51.6  & $50.6 - 53.1$ & $<51.2$ & \nod \\
        &  51.9  & $50.7 - 53.7$ &         &      \\
020819  &  51.3  & $50.4 - 52.6$ & $<51.8$ & \nod \\
        &  51.8  & $50.6 - 53.6$ &         &      \\
021004  &  51.8  & $51.4 - 52.3$ & 50.9    & \nod \\
        &  52.8  & $51.7 - 54.5$ &         &      \\
031203  &  51.1  & $50.2 - 52.5$ & 49.5    & 49.2 \\
        &  51.5  & $50.3 - 52.5$ &         &      \\
070125  &  52.2  & $51.6 - 53.2$ & 52.4    & 51.2 \\
        &  52.6  & $51.7 - 54.0$ &         &      \\
070612A &  52.1  & $51.6 - 53.2$ & $<52.0$ & \nod 
\enddata
\tablecomments{$^a$ This is the $90\%$ confidence range for the energy
of each burst.\\ 
$^b$ Values for $E_\gamma$ are taken from \citet{fks+01},
\citet{bfk03}, and \citet{fb05}.\\
$^c$ Values for $E_K$ are taken from \citet{pk02}, \citet{bkp+03},
\citet{yhs+03}, \citet{skb+04}, \citet{skb+04b}, \citet{skn+06},
\citet{cfh+10}, and \citet{cfh+10b}.\\
$^d$ The first line is for a strict cut-off of $\beta<1$, while the
second line allows a small fraction of solution with $\beta$ slightly
larger than 1 (Figure~\ref{fig:hist}). }
\end{deluxetable}

\clearpage 

\clearpage

\clearpage

\clearpage

\clearpage

\end{document}